\documentclass[aps,prb,twocolumn,showpacs,superscriptaddress,floatfix]{revtex4-1}
\usepackage{dcolumn}
\usepackage{amsmath}
\usepackage{graphicx,epsfig,psfrag}
\usepackage{amssymb}
\usepackage{natbib}
\usepackage{url}
\usepackage[breaklinks=true]{hyperref}
\usepackage{mathtools}
\usepackage{subfigure}
\hypersetup{
        colorlinks=true,
}
\usepackage{bookmark}
\usepackage{epic}

\usepackage{txfonts}
\usepackage{color}
\begin{document}
\title{Generalization of the Time-Dependent Numerical Renormalization
  Group Method to Finite Temperatures and General Pulses}
\author{H. T. M. Nghiem}
\affiliation
{Peter Gr\"{u}nberg Institut and Institute for Advanced Simulation, 
Research Centre J\"ulich, 52425 J\"ulich, Germany}
\author{T. A. Costi}
\affiliation
{Peter Gr\"{u}nberg Institut and Institute for Advanced Simulation, 
Research Centre J\"ulich, 52425 J\"ulich, Germany}
\begin{abstract}
The time-dependent numerical renormalization group (TDNRG) method [Anders {\em et al.} Phys. Rev. Lett. {\bf 95}, 196801 (2005)] offers the prospect of
investigating in a non-perturbative manner the time-dependence of local observables of interacting quantum impurity models at all time scales following
a quantum quench. Here, we present a generalization of this method to arbitrary finite temperature by making use of the full density matrix approach [Weichselbaum {\em et al.} Phys. Rev. Lett. {\bf 99}, 076402 (2007)]. We show that all terms in 
the projected full density matrix $\rho^{i\to f} = \rho^{++} + \rho^{--} + \rho^{+-}+\rho^{-+}$ appearing in the time-evolution of a local observable may 
be evaluated in closed form at finite temperature, with $\rho^{+-}=\rho^{-+}=0$. The expression for $\rho^{--}$ is shown to be finite at finite temperature, 
becoming negligible only in the limit of vanishing temperatures. We prove that this approach recovers the short-time 
limit for the expectation value of a local observable exactly at arbitrary temperatures. 
In contrast, the corresponding long-time limit is recovered exactly only for 
a continuous bath, i.e. when the logarithmic discretization parameter $\Lambda\rightarrow 1^{+}$. Since the numerical renormalization group approach breaks down in this limit, and calculations have to be carried out at $\Lambda>1$, 
the long-time behavior following an arbitrary quantum quench has a finite error, which poses an obstacle for the method, 
e.g., in its application to the scattering states numerical renormalization group method for describing steady state non-equilibrium transport through correlated impurities [Anders, Phys. Rev. Lett. {\bf 101}, 066804, (2008)]. 
We suggest a way to overcome this problem by noting that the time-dependence, in general, and 
the long-time limit, in particular, becomes increasingly more
accurate on reducing the size of the quantum quench. This suggests an improved generalized TDNRG approach in which the system is time-evolved
between the initial and final states via a sequence of small quantum quenches within a finite time interval instead of by a single large and
instantaneous quantum quench. The formalism for this is provided, thus generalizing the TDNRG method to multiple quantum quenches, periodic
switching, and general pulses. This formalism, like our finite temperature generalization of the single-quench case,  
rests on no other approximation than the NRG approximation. The results are illustrated numerically by application to the 
Anderson impurity model. 
\end{abstract}

\date{\today}

\maketitle

\section{Introduction} 
\label{sec:introduction}
The numerical renormalization group (NRG) \cite{Wilson1975,KWW1980a,KWW1980b,Bulla2008} 
has proven to be one of the most powerful methods for dealing with equilibrium properties of
strongly correlated quantum impurity systems,\cite{Hewson1997} allowing the calculation at arbitrary
temperatures, and in a non-perturbative manner, of thermodynamic,\cite{Wilson1975,KWW1980a,KWW1980b,Oliveira1981,Gonzalez-Buxton1998,Merker2012a} 
dynamic,\cite{Frota1986,Sakai1989,Costi1992,Bulla1998,Hofstetter2000,
Anders2005,Peters2006,Weichselbaum2007} and, linear transport properties\cite{Costi1994,Misiorny2012,Hanl2013} 
of such systems. The method is also applicable, within dynamical mean field theory (DMFT), \cite{Metzner1989,Georges1996,Kotliar2004,Vollhardt2012}
to correlated lattice models, such as the Kondo lattice,\cite{Costi2002,Bodensiek2013} Anderson lattice\cite{Pruschke2000} or 
Hubbard\cite{Bulla1999,Schneider2008} and Hubbard-Holstein models.\cite{Bauer2010} The introduction of the correlation self-energy, 
\cite{Bulla1998} the reduced density matrix, \cite{Hofstetter2000} the complete basis 
set of eliminated states, \cite{Anders2005} novel non-Abelian symmetries,\cite{Moca2012,Weichselbaum2012b} 
and new discretization schemes \cite{Oliveira1994,Campo2005,Zitko2009b,Mitchell2013}
have significantly improved NRG calculations, particularly for dynamic\cite{Weichselbaum2007, Peters2006,Toth2008} 
and transport properties of single and multi-channel models.\cite{Yoshida2009,Costi2010,Hanl2013,Mitchell2013} 

A further application of the method, based on the complete basis set, in combination with the reduced density matrix, 
is to the time-dependent transient response of quantum impurity systems following a quantum quench. \cite{Anders2005} 
The use of the complete basis set is particularly important here, as it resolves the problem of summing up multiple shell contributions 
to transient quantities encountered in a previous related approach. \cite{Costi1997} This time-dependent numerical renormalization
group (TDNRG) method has been used to study the transient dynamics of a number of quantum impurity models, including
the Anderson, Kondo, interacting resonant level and spin-boson models. \cite{Anders2005,Anders2006,Anders2008c,Tornow2008,Hackl2009,Orth2010}
In addition, within the scattering states NRG approach,\cite{Anders2008,Schmitt2011}
the TDNRG approach offers the prospect of investigating another important class of problems, namely 
truly non-equilibrium steady state transport through correlated nanostructures, such as through
correlated quantum dots at finite bias voltage far from the linear response regime. 
Here, the use of the TDNRG to evolve the density matrix to long times, a key ingredient 
in the approach of Ref.~\onlinecite{Anders2008}, is hampered by difficulties in obtaining 
the correct long-time limit of observables, including that of the density matrix, 
e.g., for the Anderson impurity model. \cite{Anders2005,Anders2006,Anders2008b}
In addition, there is significant noise at intermediate to long times.\cite{Anders2005,Anders2006,Anders2008b} 
It has been argued that the logarithmically discretized Wilson chain, whose heat capacity is non-extensive for discretization parameters $\Lambda>1$, 
prohibits thermalization to the correct long-time limit and possibly affects also the short-time limit.\cite{Rosch2012} 
This would pose a severe limitation on the TDNRG and its application to steady-state non-equilibrium situations. \footnote{In fact, we show
in this paper that the short-time limit is always recovered exactly by the TDNRG}
Reflections, associated with the non-constant hoppings along the Wilson chain, have also been argued to affect the long-time limit.\cite{Schmitteckert2010,Eidelstein2012} This has lead to the formulation of hybrid methods \cite{Eidelstein2012,Guettge2013} 
which combine the merits of the TDNRG method at short times, with, for example, methods, such as the time-dependent density matrix 
renormalization group method (DMRG) \cite{Cazalilla2002,Luo2003,Daley2004,White2004}, on tight-binding chains, to extract the evolution on longer time scales.

In this paper, we re-examine the TDNRG method, generalizing it to arbitrary finite temperature, for a general initial state
ensemble, within the full density matrix (FDM) approach.\cite{Weichselbaum2007} We prove a number of exact results, including
the exactness of the short-time limit of local observables following a quantum quench, and a trace conserving property of the
projected full density matrix. These prove useful in testing the finite temperature formalism. 
By applying this approach to the prototype model of strongly correlated electron systems, the Anderson impurity model, we analyze in detail, 
and shed further light on, the origin of noise in the TDNRG method and give a detailed description of errors in the local occupation and
double occupancy in the long-time limit, identifying trends in switching protocols, bath discretization and quench size that minimize these
errors. The results suggest that the short, intermediate and particularly the long-time behavior can be improved by replacing a large quantum quench 
by a sequence of smaller quantum quenches, while retaining the standard Wilson chain for the bath. With this motivation, the formalism for 
such multiple quenches is derived and shown to rest solely on the NRG approximation [Eq.~(\ref{eq:nrg-approx})], 
as in our finite temperature generalization of the single quench case. This formalism
is important since, (a), it points a way to an overall improved time-dependence within a generalized multiple-quench TDNRG approach, 
(b), it allows a general pulse (appropriately discretized, if continuous) to be treated, and, (c), it includes
periodic switching as a special case. The last, (c), for square pulses, has also been considered in Ref.~\onlinecite{Eidelstein2012}, 
where, however, additional approximations, beyond the NRG approximation, were made, and without use of the FDM 
(for a discussion of this, see the end of Sec.~\ref{sec:generic}).

%
%
%
Besides the NRG, a large number of other methods are 
being used to investigate real-time and non-equilibrium dynamics of correlated systems. 
These include analytic approaches, such as the functional renormalization group,\cite{Metzner2012} real-time renormalization group, 
\cite{Schoeller2009}  perturbative scaling approach,\cite{Rosch2003a} Keldysh perturbation theory,\cite{Kamenev2011} real-time\cite{Saptsov2012} 
and renormalized perturbation theory,\cite{Bauer2005} flow-equation, \cite{Lobaskin2005,Moeckel2008,Wang2010}, dual-fermion\cite{Jung2012}, slave-boson,\cite{Langreth1991} 
and $1/N$-expansion techniques\cite{Merino1998,Ratiani2009}. Applications of these to a number
quantum impurity models have been made, including, to the anisotropic Kondo and spin-boson models, \cite{Pletyukhov2010,Kennes2013b} to the interacting resonant level model \cite{Andergassen2011,Karrasch2010,Kennes2012a,Kennes2012b}, 
and to the Anderson impurity model\cite{Meir1993,Shao1994a,Shao1994b,Goker2007,Oguri2001,Bauer2005,Munoz2013,Ratiani2009,Saptsov2012,Saptsov2013}. 
One advantage of these analytic approaches over the TDNRG, and other
numerical approaches, is their ability to deal with a continuous bath. On the other hand, most of these approaches are perturbative, and they may have difficulty 
accessing the low temperature strong coupling limit. 
Numerical methods include the aforementioned DMRG method\cite{Daley2004,White2004}, also at finite temperature,\cite{Feiguin2005,Barthel2009,Karrasch2013} 
and several quantum Monte Carlo approaches.\cite{Cohen2013,Cohen2013b,Gull2011,Weiss2008b,Muehlbacher2008} 
The former is limited, as compared to TDNRG, to finite size systems and is not applicable to exponentially long times,\footnote{The Kondo model, for example, exhibits interesting physics at exponentially long times, $t\gg \hbar/T_{\rm K}$, where $T_{\rm K}\sim \sqrt{JN_{\rm F}}e^{-1/JN_{\rm F}}$ is the
exponentially small Kondo scale, with $J$ the exchange constant and $N_{\rm F}$ the density of states} whereas the latter become
computationally expensive for strong interactions or low temperatures. Indeed, the main motivation for 
developing the TDNRG approach, is, on the one hand, its inherently non-perturbative nature, allowing strongly correlated systems
to be treated accurately at moderate computational cost,  with, on the other hand, 
its ability to describe arbitrary low energy scales and temperatures. The latter, in principle, 
has the prospect of addressing accurately the time-dependence of such systems at arbitrarily long times. 
These advantages, however, come at the expense of using a logarithmically discretized bath, which incurs errors that we shall address in this paper. 
Finally, we mention that most of the above techniques, are, in principle, also of use in the description of non-equilibrium DMFT for
correlated lattice models. \cite{Freericks2006,Aoki2013,Arrigoni2013} 
%
%

The present work is also motivated by the increasing number of experiments probing time-dependent properties of correlated systems, 
including time-dependent spectroscopies of correlated electron systems, such as pump-probe investigations of Mott insulators, 
\cite{Perfetti2006,Perfetti2008} quasiparticle lifetime effects in surface states,\cite{Loukakos2007} coherent control and relaxation times 
in solid-state qubits, \cite{Petta2005,Koppens2006} determination of relaxation rates of excited spin states of atoms on surfaces via voltage pulses, 
\cite{Loth2010} and non-equilibrium effects in cold atom systems. \cite{Greiner2002,Will2010,Trotzky2012,Schneider2012}

The paper is organized as follows. In Sec.~\ref{sec:section-1}, we provide required background information (Sec.~\ref{subsec:qi+qq}-\ref{subsec:time-dependence-observables}), 
discuss limiting cases and exact results (Sec.~\ref{subsec:exact results}), and
state the problem that has to be overcome in generalizing the existing TDNRG approach of Refs.~\onlinecite{Anders2005,Anders2006} to
finite temperature within the FDM formalism (Sec.~\ref{subsec:generalizing-tdNRG}).
This generalization requires calculating all terms appearing in a projected density matrix, which 
is accomplished in Sec.~\ref{sec:section-2}. In Sec.~\ref{sec:section-3} we derive recursion relations for these terms, 
which allow them to be calculated in a numerically efficient manner.
We then apply this finite-temperature generalization of the TDNRG approach to the Anderson impurity model
in Sec.~\ref{sec:section-4}, providing detailed numerical results for local observables on all time and temperature scales. In particular we
discuss the accuracy of this approach for the occupation number and double occupancy at different temperature and time scales.
In Sec.~\ref{sec:generic}, we generalize the TDNRG approach from a single instantaneous quantum quench to a sequence of multiple quantum 
quenches over a finite time interval. The primary motivation for this generalization was stated above, namely to obtain an overall
improvement in the accuracy of the TDNRG method at longer times. However, in the process we generalize 
the approach to other cases of great interest in their own right, such as to periodic switching/driving and 
to general pulses (suitably discretized, if continuous). A summary and outlook on future prospects and applications of our formalism
is presented in Sec.~\ref{sec:conclusions}. The trace conserving property of the projected full density matrix is shown in Appendix~\ref{sec:trace-formula}, and Appendix~\ref{sec:appendix-short-time-limit} gives a detailed
proof that the short-time limit of observables is exact within TDNRG. 
Appendix~\ref{sec:su2} gives explicit expressions for recursion relations for the projected density matrix 
for the case of SU(2) spin symmetry. We used this symmetry in 
obtaining the numerical results for the Anderson impurity model in Sec.~\ref{sec:section-4}. 
Additional numerical results for the Anderson model are given in Appendix~\ref{sec:supporting}, while further details of the TDNRG derivation for multiple quenches are given in Appendix~\ref{sec:tau-overlap-matrices}.

\section{Preliminaries}
\label{sec:section-1}
The NRG and its extension to time-dependent problems apply generally to any quantum impurity model. For the purposes of this
section, we outline these,  together with the generic quantum quench of interest in Sec.~\ref{subsec:qi+qq}, postponing a detailed
description of the specific model that we shall use later in the numerical calculations (the Anderson impurity model) and the specific 
quench protocols to Sec.~\ref{sec:section-4}. Some general aspects of the NRG together with the 
complete basis formalism\cite{Anders2005,Anders2006}, and the FDM\cite{Weichselbaum2007} 
are introduced in Sec.~\ref{subsec:nrg+cbs}.  We recapitulate the use of the complete basis set to derive a formal
expression for the time-dependence of a local observable following a quantum quench in Sec.~\ref{subsec:time-dependence-observables}. 
Limiting cases and exact results that hold generally within the time-dependent NRG are given in Sec.~\ref{subsec:exact results}. Finally, 
Sec.~\ref{subsec:generalizing-tdNRG} states the problem to be overcome in generalizing the approach of Refs.~\onlinecite{Anders2005,Anders2006} 
to finite temperature, using the FDM,\cite{Weichselbaum2007} the main aim of the present paper. 
\subsection{Quantum impurity models and quantum quenches}
\label{subsec:qi+qq}
The Hamiltonian of a quantum impurity system is given by
\begin{align}
&H = H_{\rm imp} + H_{\rm int} + H_{\rm bath},\label{eq:QIS}
\end{align}
where $H_{\rm imp}$ represents the impurity, a small quantum mechanical system with a few degrees of freedom and local many-body interactions, $H_{\rm bath}$ represents a quasi-continuous bath of non-interacting particles, and $H_{\rm int}$ is the interaction between the two. Typical examples include the Anderson impurity model (Sec.~\ref{subsec:model}), the anisotropic Kondo model, or the spin-boson model of two-level systems coupled to an environment.\cite{Leggett1987,Weiss2008,LeHur2008} 
We are interested in the dynamics of a local observable $\hat{O}$ following a quantum quench in which one or more system parameters of $H$ change suddenly at $t=0$. Thus, the time dependence of $H$ is described by
$H(t)=\theta(-t)H^{i} + \theta(t)H^{f}$, with  $H^{i}$ and $H^{f}$ being time-independent 
initial ($t<0$) and final state ($t>0$) Hamiltonians, respectively. 
The time evolution of $\hat{O}$ at $t>0$ is then given by $O(t)= {\rm Tr}\left[\rho(t)\hat {O}\right]$ where $\rho(t)=e^{-iH^{f}t}\rho\,e^{iH^{f}t}$ is the time-evolved density matrix and
$\rho=e^{-\beta H^{i}}/{\rm Tr}[\rho]$ is the  density matrix of the initial state at inverse temperature $\beta$. 
\subsection{NRG, complete basis set, and full density matrix}
\label{subsec:nrg+cbs}
Since  the initial and final state Hamiltonians $H^{i}$ and $H^{f}$ are time-independent they can  be iteratively diagonalized in the usual way within the NRG method. 
In brief, this approach, described in detail in Refs.~\onlinecite{Wilson1975,KWW1980a,KWW1980b,Bulla2008}, yields the eigenstates and eigenvalues of a sequence of truncated Hamiltonians $H_{m}^{i/f}, m=1,2,\dots$, which approximate the spectra of $H^{i/f}$, on successively decreasing energy
scales $\omega_{m}\sim \Lambda^{-m/2}$. The discretization parameter $\Lambda>1$ is
required to achieve a separation of energy scales in $H^{i/f}$, such that an iterative diagonalization scheme can be successfully carried out. This procedure is performed up to a maximum iteration $m=N$ (``the longest Wilson chain'').  At each $m$, the states generated, denoted $|qm\rangle$,
are partitioned into the lowest energy retained states, denoted $|km\rangle$,  and the high energy eliminated (or discarded) states, $| lm\rangle$. In order to avoid an exponential increase
in the dimension of the Hilbert space, only the former are used to set up and diagonalize the Hamiltonian for the next iteration $m+1$. The eliminated states, 
while not used in the iterative NRG procedure, are nevertheless crucial as they are used to set up a complete basis set with which the expressions for the time dependent dynamics 
are evaluated.\cite{Anders2005} This complete basis set is defined by the product states $|lem\rangle=|lm\rangle |e\rangle, m=m_{0},\dots,N$, where $m_{0}$ is the first iteration at which truncation occurs,
and $|e\rangle=|\alpha_{m+1}\rangle |\alpha_{m+2}\rangle\dots|\alpha_{N}\rangle$ are environment states at iteration $m$ such that the product states $| lem\rangle$, for each $m=m_{0},m_{0}+1,\dots,N$, 
reside in the same Fock space (that of the largest system diagonalized, $m=N$). The $\alpha_m$ represent the configurations of site $m$ in a linear chain representation of the quantum impurity system 
(e.g. the four states $0$, $\uparrow$, $\downarrow$ and $\uparrow\downarrow$ at site $m$ for a single channel Anderson model) and ``$e$'' in $|lem\rangle$ denotes the
collection $e=\{\alpha_{m+1}\dots\alpha_N\}$.  These states satisfy the completeness relation \cite{Anders2005,Anders2006}
\begin{align}
&\sum_{m=m_0}^N \sum_{le} | lem\rangle\langle lem|=1,\label{unity-decomposition}
\end{align}
where for $m=N$ all states are counted as discarded (i.e. there are no kept states at iteration $m=N$).
We shall also use the following representations of this relation, \cite{Anders2005,Anders2006}
\begin{align} 
&1=1^{-}_{m}+1^{+}_{m},\label{eq:unity-decomposition-a}\\
&1_{m}^{-}=\sum_{m'=m_{0}}^{m}\sum_{l'e'}|l'e'm'\rangle\langle l'e'm'|\label{eq:unity-decomposition-b}\\
&1_{m}^{+}=\sum_{m'=m+1}^{N}\sum_{l'e'}|l'e'm'\rangle\langle l'e'm'| =\sum_{ke}|kem\rangle\langle\,kem|. \label{eq:unity-decomposition-c}
\end{align}

By using the complete basis set, we can construct an initial state density matrix which is valid at any temperature, the 
FDM, \cite{Weichselbaum2007,Weichselbaum2012} 
\begin{align}
&\rho=\sum_{m=m_0}^N w_m \tilde{\rho}_m,\label{eq:fdm-initial-state}\\
&\tilde{\rho}_m=\sum_{le}|lem\rangle{_i} \frac{e^{-\beta E_l^m}}{\tilde{Z}_m}{_i}\langle lem|,
\end{align}
which takes into account the discarded states of $H^{i}$ from all shells. For later use, we note that, 
(a), ${\rm Tr}\left[\tilde{\rho}_{m}\right]={\rm Tr}\left[\rho\right]=1$ implies that $\sum_{m=m_0}^{N}w_m =1$, and, 
(b),  ${\rm Tr}\left[\tilde{\rho}_m\right]=1$ implies that $1=\sum_{le}\frac{e^{-\beta E_{l}^m}}{\tilde{Z}_{m}}=\sum_{l}d^{N-m}\frac{e^{-\beta E_{l}^m}}{\tilde{Z}_{m}}=d^{N-m}\frac{Z_{m}}{
\tilde{Z}_{m}}$ where $Z_{m}=\sum_{l}e^{-\beta E_{l}^{m}}$, i.e., $\tilde{Z}_m = d^{N-m}Z_m$, and $d$ is the degeneracy of the Wilson site $\alpha_m$. \cite{Weichselbaum2007,Costi2010} 

\subsection{Time-dependence}
\label{subsec:time-dependence-observables}
With the above definitions, we can write for the time evolution of a local observable $\hat{O}$ 
\begin{align}
&O(t) = {\rm Tr}\left[e^{-iH^{f}t}\rho e^{iH^{f}t}\hat{O}\right]\nonumber\\
&=\sum_{m=m_0}^N \sum_{le} {_f}\langle lem|e^{-iH^f t}\rho e^{iH^f t}\hat{O}| lem\rangle_f\nonumber\\
&=\sum_{mm'=m_0}^N \sum_{lel'e'} {_f}\langle lem|e^{-iH^f t}\rho e^{iH^f t}| l'e'm'\rangle_f  {_f}\langle l'e'm'| \hat{O}| lem\rangle_f\label{eq:Ot}.
\end{align}
Here, and in the remainder of the paper, $A_{i/f}$ denotes an expression in the basis of initial/final states. 
Using $\sum_{ke}| kem\rangle \langle kem|=\sum_{m'=m+1}^N \sum_{l'e'}| l'e'm'\rangle \langle l'e'm'|$ [i.e., the second equality in Eq.~(\ref{eq:unity-decomposition-c})], we
can express Eq.~({\ref{eq:Ot}}) as\cite{Anders2005,Anders2006}
\begin{align}
O(t) &=\sum_{m=m_0}^N \sum_{rs\notin KK'} \sum_e{_f}\langle sem|e^{-iH^f t}\rho e^{iH^f t} | rem\rangle_f {_f}\langle rem| \hat{O}| sem\rangle_f\nonumber\\
&\approx\sum_{m=m_0}^N \sum_{rs\notin KK'} \sum_e{_f}\langle sem|e^{-iH^f_m t}\rho e^{iH^f_m t} | rem\rangle_f {_f}\langle rem| \hat{O}| sem\rangle_f\nonumber\\
&=\sum_{m=m_0}^N \sum_{rs\notin KK'} \Big(\sum_e{_f}\langle sem|\rho | rem\rangle_f\Big) e^{-i(E^m_s-E^m_{r}) t} O^m_{rs} \nonumber\\
&=\sum_{m=m_0}^N \sum_{rs\notin KK'}\rho^{i\to f}_{sr}(m) e^{-i(E^m_s-E^m_{r}) t} O^m_{rs} \label{eq:localOt},
\end{align}
in which $r$ and $s$ may not both be kept states, $O^m_{rs}={_f}\langle lem|\hat{O}|rem\rangle_f$ are the final state matrix elements of $\hat{O}$,
which are independent of $e$, the NRG approximation 
\begin{align}
&H^{f}|rem\rangle\approx H^{f}_{m}|rem\rangle=E^m_r|rem\rangle,\label{eq:nrg-approx}
\end{align}
is adopted [in the second line of Eq.~(\ref{eq:localOt})], and $\rho^{i\to f}_{sr}(m)=\sum_e{_f}\langle sem|\rho | rem\rangle_f$ represents the reduced density matrix of the initial state projected onto the basis of final states (henceforth simply called the
{\em projected density matrix}). 
\subsection{Limiting cases and exact results}
\label{subsec:exact results}
A number of limiting cases and exact results follow generally from the above formalism. First, it is clear from Eq.~(\ref{eq:localOt}) 
that in the absence of a quench, $H^i = H^f$, one recovers for $O(t)$ the time independent equilibrium thermodynamic average 
\begin{align}
&O(t)=O_i = O_f.\label{eq:equilibrium-limit}
\end{align}
Second, in the special case that $\hat{O}$ is the identity operator, we have from Eq.~(\ref{eq:localOt}) upon using $O^m_{rs}=\delta_{rs}$ that
\begin{align}
1=\sum_{m=m_0}^N \sum_{l}\rho^{i\to f}_{ll}(m)\label{eq:identity},
\end{align}
i.e., the trace of the projected density matrix over the eliminated states is conserved (see Appendix~\ref{sec:trace-formula}).
The above two results will serve as useful checks on the correctness of our finite temperature generalization of the time dependent 
NRG as well as on the precision of the numerical calculations in Sec.~\ref{sec:section-4}.

Third, it has been noted previously that the TDNRG yields very accurate results for the short-time limit of observables, i.e., to a very good
approximation $O(t\rightarrow 0^{+})\approx O_i$ where  $O_i = {\rm Tr}\left[\rho\hat{O}\right]$ is the thermodynamic value in the initial state.\cite{Anders2006} In fact, we can show that 
the short-time limit is exact, i.e., $O(t\rightarrow 0^{+})= O_i$.  An explicit proof of this, starting from the expression for $O(t\to 0^{+})$,
is given in Appendix~\ref{sec:appendix-short-time-limit}. One can also see this from the following argument: the time evolution in Eq.~(\ref{eq:localOt}) involves 
approximate NRG eigenvalues via the factor $e^{i(E_r^m - E_s^m)t}$. The NRG approximation for the eigenvalues
is therefore not operative for $t\to 0^+$ and Eq.~(\ref{eq:localOt}) is as exact as  Eq.~(\ref{eq:Ot}) in this limit (the latter yielding $O_i={\rm Tr}[\rho \hat{O}]$), hence
we have the exact result,\footnote{By {\em exact} we mean the equality stated in Eq.~(\ref{eq:Ot0}). The actual value evaluated with NRG may deviate by a small
amount from exact analytic results, e.g. from Bethe ansatz results, however, this difference has been shown to be negligible \cite{Merker2012a, Merker2012b}}
\begin{align}
O(t\to 0^{+})\equiv\sum_{m=m_0}^N \sum_{rs\notin KK'}\rho^{i\to f}_{sr}(m) O^m_{rs}= O_i \label{eq:Ot0}.
\end{align}
This will also be verified numerically in Sec.~\ref{sec:section-4}. 

In contrast, as soon as $t$ is finite, we expect that the time evolution in  Eq.~(\ref{eq:localOt})  will differ from
that in Eq.~(\ref{eq:Ot}), and this will, in part, be due to the appearance of approximate NRG eigenvalues in the former, resulting
in errors and noise in the time evolution, which we shall analyze numerically in Sec.~\ref{subsec:time-dependence}.
We thus also expect, and find, that the long-time limit of $O(t)$, given by
\begin{align}
&O(t\to \infty)=\sum_{m=m_0}^N \sum_{rs\in DD'}\rho^{i\to f}_{sr}(m) \delta_{E^m_s,E^m_{r}} O^m_{rs}\label{eq:localOtInfinity},\\
& = \sum_{m=m_0}^N \sum_{l}\rho^{i\to f}_{ll}(m)O^m_{ll},\label{eq:localOtInf}
\end{align}
is not exact. The deviation of this from its expected thermodynamic value in the final state, $O_f$, will be extensively 
analyzed in Sec.~\ref{subsec:long-time-limit}.

\subsection{Generalizing the time-dependent NRG to finite temperature}
\label{subsec:generalizing-tdNRG}
We now return to the formal expression for $O(t)$ in Eq.~(\ref{eq:localOt}) and consider its explicit evaluation. 
The matrix elements $O^m_{rs}$ in Eq.~(\ref{eq:localOt}) may be calculated recursively at each shell, in the usual way within the NRG.\cite{KWW1980a,Bulla2008} 
Calculating the projected density matrix $\rho^{i\to f}_{sr}(m)$ requires more effort and is the main problem in generalizing the TDNRG to finite temperatures. 
In order to see this more clearly, we use a modification of Eqs.~(\ref{eq:unity-decomposition-a})-(\ref{eq:unity-decomposition-c}) for the decomposition of unity as in Ref.~\onlinecite{Anders2006} 
\begin{align}
1&=\mathcal{I}^+_m+\mathcal{I}^-_m \nonumber\\
&=\sum_{qe} | qem\rangle_i {_i}\langle qem|+\sum_{m'=m_0}^{m-1}\sum_{l'e'} | l'e'm'\rangle_i {_i}\langle l'e'm'|,
\end{align}
in which the label $q$ includes both retained and eliminated states at iteration $m$, and break $\rho^{i\to f}_{sr}(m)$ into four terms
\begin{align}
\rho^{i\to f}_{sr}(m)&=\sum_e{_f}\langle sem|(\mathcal{I}^+_m+\mathcal{I}^-_m)\rho(\mathcal{I}^+_m+\mathcal{I}^-_m) | rem\rangle_f\nonumber\\
&=\rho^{++}_{sr}(m)+\rho^{+-}_{sr}(m)+\rho^{-+}_{sr}(m)+\rho^{--}_{sr}(m)\label{eq:totalrho},
\end{align}
with $\rho^{pp'}_{sr}(m)=\sum_e{_f}\langle sem|\mathcal{I}^p_{m}\rho \mathcal{I}^{p'}_{m} | rem\rangle_f$ and $p,p'=\pm$. Calculating $\rho^{++}_{sr}(m)$ is straightforward, since it includes only overlap matrix elements 
between initial and final states at the same shell, ${_i}\langle qem|rem\rangle_f$. But for $\rho^{+-}_{sr}(m)$, $\rho^{-+}_{sr}(m)$, and $\rho^{--}_{sr}(m)$ this is manifestly not the case, requiring
overlap matrix elements between initial and final states at different shells, ${_i}\langle l'e'm'| rem\rangle_f$ with $m'<m$.

In previous work,\cite{Anders2006} the problem of evaluating $\rho^{+-}_{sr}(m)$, $\rho^{-+}_{sr}(m)$, and $\rho^{--}_{sr}(m)$, for general situations,
was avoided by choosing a special form for the initial state density matrix, namely  
\begin{align}
&\rho=\sum_{l}|lN\rangle{_i} \frac{e^{-\beta E_l^N}}{Z_N}{_i}\langle lN|, \label{eq:rho-Anders}
\end{align}
with $Z_{N}=\sum_{l}e^{-\beta E_{l}^{N}}$, in which only the discarded states of the last NRG iteration enter, while discarded states at $m<N$ are neglected.  
This simplifies the projected density matrix to $\rho^{i\to f}_{sr}(m)=\rho^{++}_{sr}(m)$ as  $\rho^{+-}_{sr}(m)=\rho^{-+}_{sr}(m)=\rho^{--}_{sr}(m)=0$ for all $m\le N$ for
this special choice of $\rho$. However, this limits the calculations to $T\sim T_N$ where $T_N \sim \omega_N$ is the characteristic scale of $H_{N}^{i}$.
Calculations at higher temperature $T>T_N$, within this approach, would require choosing a shorter chain of length $M<N$ such that $T\approx T_M$,
which introduces additional errors due to the shorter chain. Hence, a formulation that uses the FDM of the initial state [Eq.~(\ref{eq:fdm-initial-state})], 
which encodes all discarded states from
all shells $m\le N$, would be advantageous, since it would be valid at all temperatures $T\ge T_N$ automatically, while a fixed chain of length $N$ is used for all $T$.
In the next section, we show that this is possible, hence allowing the calculation of the time-dependence at an arbitrary finite temperature starting from a general initial
state. We shall thus use the FDM in Eq.~(\ref{eq:fdm-initial-state}), which takes into account the discarded states from all shells, and present the calculation of 
each of the four terms appearing in $\rho^{i\to f}_{sr}(m)$, especially $\rho^{+-}_{sr}(m)$, $\rho^{-+}_{sr}(m)$, and $\rho^{--}_{sr}(m)$. 

\section{Finite-temperature projected density matrix}
\label{sec:section-2}
In this section we evaluate the four terms contributing to
$\rho^{i\to f}_{sr}(m)$ in Eq.~(\ref{eq:totalrho}) starting from the
FDM of the initial state $\rho=\sum_{m=m_0}^N w_m \tilde{\rho}_m$. 
Substituting the latter into the first term of $\rho^{i\to f}_{sr}(m)$ in Eq.~(\ref{eq:totalrho}), we have
\begin{align}
&\rho^{++}_{sr}(m)=\sum_{m'=m_0}^N w_{m'}\sum_e{_f}\langle sem|\mathcal{I}^+_m\tilde{\rho}_{m'}  \mathcal{I}^+_m| rem\rangle_f\nonumber\\
&=\sum_{m'=m_0}^N \sum_{\substack{eqe'\\q'e''}}{_f}\langle sem| qe'm\rangle_i {_i}\langle qe'm|w_{m'}\tilde{\rho}_{m'} |q'e''m\rangle_i {_i}\langle q'e''m| rem\rangle_f\nonumber,
\end{align}
in which the overlap matrix elements, ${_i}\langle qe'm|
rem\rangle_f=S_{q_ir_f}^{m}\delta_{ee'}$, between initial and final
states, are diagonal in, and independent of the environment degrees of freedom. Then
\begin{align}
&\rho^{++}_{sr}(m)=\sum_{m'=m_0}^N \sum_{qq'}S_{s_fq_i}^{m} \sum_{e}{_i}\langle qem|w_{m'}\tilde{\rho}_{m'} |q'em\rangle_i S_{q_ir_f}^{m}\nonumber\\
&=\sum_{m'=m_0}^N \sum_{qq'}S_{s_fq_i}^{m} \sum_{e,le'}{_i}\langle qem|le'm'\rangle{_i} w_{m'}\frac{e^{-\beta E_l^{m'}}}{\tilde{Z}_{m'}}{_i}\langle le'm' |q'em\rangle_i S_{q'_ir_f}^{m}\nonumber.
\end{align}
Since $\sum_q|qem\rangle=\sum_l|lem\rangle+\sum_k|kem\rangle$, ${_i}\langle l'e'm' |lem\rangle_i=\delta_{ll'}\delta_{ee'}\delta_{mm'}$, and ${_i}\langle le'm' |kem\rangle_i = 0$ as $m'\leq m$, the above equation is equivalent to
\begin{align}
&\rho^{++}_{sr}(m)=\sum_{le}S_{s_fl_i}^{m}w_{m} \frac{e^{-\beta E_l^{m}}}{\tilde{Z}_{m}} S_{l_ir_f}^{m}\nonumber\\
&+\sum_{m'=m+1}^N \sum_{kk'}S_{s_fk_i}^{m} \sum_{ele'}{_i}\langle kem|le'm'\rangle{_i} w_{m'}\frac{e^{-\beta E_l^{m'}}}{\tilde{Z}_{m'}}{_i}\langle le'm' |k'em\rangle_i S_{k'_ir_f}^{m}\nonumber.
\end{align}
Using the definition of the full reduced density matrix\cite{Costi2010} 
\begin{align}
R_{\rm red}^m(k,k')&=\nonumber\\
\sum_{m'=m+1}^N&\sum_{ele'}{_i}\langle kem|le'm'\rangle{_i} w_{m'}\frac{e^{-\beta E_l^{m'}}}{\tilde{Z}_{m'}}{_i}\langle le'm' |k'em\rangle_i\label{eq:RFDM},
\end{align}
and $\tilde{Z}_m = d^{N-m}Z_m$ from Sec.~\ref{sec:section-1},  we have 
\begin{align}
\rho^{++}_{sr}(m)=\nonumber\\
\sum_{l}&S_{s_fl_i}^{m} w_{m}\frac{e^{-\beta E_l^{m}}}{Z_{m}} S_{l_ir_f}^{m}+\sum_{kk'}S_{s_fk_i}^{m}R_{\rm red}^m(k,k')S_{k'_ir_f}^{m}\label{eq:rho++}.
\end{align}

In the other terms with $\mathcal{I}^-_m$, we notice that\cite{Costi2010} 
\begin{equation}
\mathcal{I}^-_m\tilde{\rho}_{m'}=\tilde{\rho}_{m'}\mathcal{I}^-_m=\begin{dcases}
     \tilde{\rho}_{m'} & \text{if } m'<m;\\
0 & \text{otherwise}\nonumber,
   \end{dcases}
\end{equation}
then
\begin{align}
&\rho^{+-}_{sr}(m)=\sum_{m'=m_0}^N w_{m'}\sum_e{_f}\langle sem|\mathcal{I}^+_m\tilde{\rho}_{m'}  \mathcal{I}^-_m| rem\rangle_f\nonumber\\
&=\sum_{m'=m_0}^{m-1} w_{m'}\sum_{eqe'}{_f}\langle sem| qe'm\rangle_i {_i}\langle qe'm|\tilde{\rho}_{m'}| rem\rangle_f\nonumber\\
&=\sum_{m'=m_0}^{m-1} \sum_{\substack{eqe'\\le''}}{_f}\langle sem| qe'm\rangle_i {_i}\langle qe'm|le''m'\rangle{_i} w_{m'}\frac{e^{-\beta E_l^{m'}}}{\tilde{Z}_{m'}}{_i}\langle le''m'| rem\rangle_f\nonumber.
\end{align}
This term equals zero since ${_i}\langle qe'm|le''m'\rangle{_i}=0$
with $m'<m$ (i.e., discarded states $|le''m'\rangle$ at iteration
$m'<m$ have no overlap with states $|qe'm\rangle$ of later
iterations). 
Similarly, $\rho^{-+}_{sr}(m)=0$. 

Finally, the last term contributing to $\rho_{sr}^{i\to f}(m)$ is given by
\begin{align}
\rho^{--}_{sr}(m)=&\sum_{m'=m_0}^N w_{m'}\sum_e{_f}\langle sem|\mathcal{I}^-_m\tilde{\rho}_{m'}  \mathcal{I}^-_m| rem\rangle_f\nonumber\\
=&\sum_{m'=m_0}^{m-1}\sum_{e,le'}{_f}\langle sem|le'm'\rangle{_i}  w_{m'}\frac{e^{-\beta E_l^{m'}}}{\tilde{Z}_{m'}}{_i}\langle le'm'| rem\rangle_f\nonumber.
\end{align}
Inserting $1=1^+_{m'}+1^-_{m'}$ from Eq.~(\ref{eq:unity-decomposition-a}) into the overlap matrix elements
appearing above, we have that
\begin{align}
&{_f}\langle sem|le'm'\rangle{_i}={_f}\langle sem|(1^+_{m'}+1^-_{m'})|le'm'\rangle{_i}\nonumber\\
&=\sum_{k} {_f}\langle sem|ke'm'\rangle{_f}{_f}\langle ke'm'|le'm'\rangle{_i}\nonumber\\
&+\sum_{m''=m_0}^{m'}\sum_{l'e''} {_f}\langle sem|l'e''m''\rangle{_f}{_f}\langle l'e''m''|le'm'\rangle{_i}\nonumber\\
&= \sum_k {_f}\langle sem|ke'm'\rangle{_f} S^{m'}_{k_fl_i}\end{align}
since ${_f}\langle sem|l'e''m''\rangle{_f}=0$ for $m''<m$, then
\begin{align}
\rho^{--}_{sr}(m)=&\sum_{m'=m_0}^{m-1}\sum_{ee'kk'}{_f}\langle sem|ke'm'\rangle{_f}\nonumber\\
&\times\Bigg[\sum_l S^{m'}_{k_{fl_i}}\Big(w_{m'}\frac{e^{-\beta E_{l}^{m'}}}{\tilde{Z}_{m'}}\Big)S^{m'}_{l_ik'_{f}}\Bigg]{_f}\langle k'e'm'| rem\rangle_f\label{eq:rho--}.
\end{align}

In general, we proved that $\rho^{+-}_{sr}(m)=\rho^{-+}_{sr}(m)=0$ and 
\begin{equation}
\rho^{i\to f}_{sr}(m)=\rho^{++}_{sr}(m)+\rho^{--}_{sr}(m),\label{eq:two-contributions}
\end{equation} 
without the restriction of using a density matrix defined for the last iteration, therefore the calculation of $O(t)$ is possible at arbitrary temperatures for a general initial state. Furthermore, in Eq.~(\ref{eq:rho--}), $\rho^{--}_{sr}(m)$ is no longer expressed in terms of overlap matrix elements ($S$-matrix elements) between initial and final states belonging to different shells but instead involves only shell diagonal matrix elements. This simplifies the calculation of $\rho^{--}_{sr}(m)$ making it as simple as the calculation of $R^m_{\rm red}(k_i,k'_i)$\cite{Weichselbaum2007}: notice the structural similarities between Eq.~(\ref{eq:RFDM}) and Eq.~(\ref{eq:rho--}). The $R^{m}_{\rm red}$ are obtained
efficiently by tracing out shell degrees of freedom backwards starting from $N$ in a recursive manner. Similarly,  the $\rho^{--}(m)$ can be obtained recursively by a forward iteration procedure starting from
$m=m_{0}$. This forward recursive procedure, described in the next section, makes the calculation of $\rho^{--}(m)$ as efficient as that for $R^{m}_{\rm red}$.
\section{Recursion relations for the projected density matrix}
\label{sec:section-3}
As shown in the last section, the projected density matrix is
simplified into two contributions [Eq.~(\ref{eq:two-contributions})]. 
In this section, we show how to calculate them recursively. For
convenience, we rewrite Eq.~(\ref{eq:two-contributions}) as
\begin{equation}
\rho^{i\to f}_{sr}(m)=\tilde{\rho}^{++}_{sr}(m)+\rho^{0}_{sr}(m)+\rho^{--}_{sr}(m)\label{eq:totalRho}, 
\end{equation}
where
\begin{align}
\tilde{\rho}^{++}_{sr}(m)&=\sum_{kk'}S_{s_fk_i}^{m}R_{\rm red}^m(k,k')S_{k'_ir_f}^{m},\label{eq:rho++b}\\
\rho^0_{sr}(m)&=\sum_{l}S_{s_fl_i}^{m} \Big(w_{m}\frac{e^{-\beta E_l^{m}}}{Z_{m}} \Big) S_{l_ir_f}^{m}\label{eq:rho0}.
\end{align}
$\rho^{0}_{sr}(m)$ can be calculated easily at each shell once we have
all the eigenvalues and overlap matrix elements. $\tilde{\rho}^{++}_{sr}(m)$
can also be calculated at each shell, once $R_{\rm red}^m(k_i,k'_i)$ has
been calculated recursively. Here, we present the recursive
calculation for $\rho^{--}_{sr}(m)$, with the calculation for $R_{\rm red}^m(k_i,k'_i)$ being similar, \cite{Weichselbaum2007,Toth2008}

Using the above definition of $\rho^{0}_{sr}(m)$ and
$\frac{1}{Z_{m'}}=\frac{1}{\tilde{Z}_{m'}}d^{N-m'}$, we have that
Eq.~(\ref{eq:rho--}) is equivalent to
\begin{align}
\rho^{--}_{sr}(m)=&\sum_{m'=m_0}^{m-1}\frac{1}{d^{N-m'}}\sum_{\substack{ee'\\kk'}}{_f}\langle sem|ke'm'\rangle{_f}\rho^{0}_{kk'}(m'){_f}\langle k'e'm'| rem\rangle_f\nonumber.
\end{align}
We note that $\rho^{--}_{sr}(m_0)=0$, and 
\begin{align}
&\rho^{--}_{sr}(m_0+1)\nonumber\\
&=\frac{1}{d^{N-{m_0}}}\sum_{\substack{ee'\\kk'}}{_f}\langle se(m_0+1)|ke'm_0\rangle{_f}\rho^{0}_{kk'}(m_0){_f}\langle k'e'm_0| re(m_0+1)\rangle_f\nonumber\\
&=\frac{1}{d}\sum_{\alpha_{m_0+1}kk'}A^{\alpha_{m_0+1}\dagger}_{sk}\rho^{0}_{kk'}(m_0)A^{\alpha_{m_0+1}}_{k'r}\label{eq:rho--m0+1}
\end{align}
since ${_f}\langle k'e'm_0|
re(m_0+1)\rangle_f=\delta_{e'e}A^{\alpha_{m_0+1}}_{k'r}$ with
$e'=\{e,\alpha_{m_0+1}\}$, and $\sum_e=d^{N-(m_0+1)}$. In the above we used the
notation of Ref.~\onlinecite{Weichselbaum2007} for the transformation matrix
$A^{\alpha_{m_0+1}}_{k'r}$ relating eigenstates $|r(m_0+1)\rangle$ of $H_{m_0+1}$ to the product basis
$|k'm_0\rangle|\alpha_{m_0+1}\rangle$,  i.e.,
\begin{equation}
|r(m_0+1)\rangle =
\sum_{k'\alpha_{m_0+1}}A^{\alpha_{m_0+1}}_{k'r}|k'm_0\rangle|\alpha_{m_0+1}\rangle.\label{eq:transformation-matrix}
\end{equation}
Similarly, we have that
\begin{align}
\rho^{--}_{sr}(m_0+2)&=\frac{1}{d}\sum_{\substack{\alpha_{m_0+2}\\kk'}}A^{\alpha_{m_0+2}\dagger}_{sk}\rho^{0}_{kk'}(m_0+1)A^{\alpha_{m_0+2}}_{k'r}\nonumber\\
&+\frac{1}{d^2}\sum_{\substack{\alpha_{m_0+2}\alpha_{m_0+1}\\kk'k_1k'_1}}A^{\alpha_{m_0+2}\dagger}_{sk}A^{\alpha_{m_0+1}\dagger}_{kk_1}\rho^{0}_{k_1k'_1}(m_0)A^{\alpha_{m_0+1}}_{k'_1k'}A^{\alpha_{m_0+2}}_{k'r}.\nonumber
\end{align}
Using the definition of $\rho^{--}_{sr}(m_0+1)$ in Eq.~(\ref{eq:rho--m0+1}), we obtain
\begin{align}
&\rho^{--}_{sr}(m_0+2)=\frac{1}{d}\sum_{\substack{\alpha_{m_0+2}\\kk'}}A^{\alpha_{m_0+2}\dagger}_{sk}[\rho^{0}_{kk'}(m_0+1)+\rho^{--}_{kk'}(m_0+1)]A^{\alpha_{m_0+2}}_{k'r}\nonumber.
\end{align}
Consequently, we have the recursion relation
\begin{equation}
\rho^{--}_{sr}(m)=\begin{dcases}
     0&\text{if }m=m_0;\\
\frac{1}{d}\sum_{\substack{\alpha_{m}\\kk'}}A^{\alpha_{m}\dagger}_{sk}\big[\rho^0_{kk'}(m-1)+\rho^{--}_{kk'}(m-1)\big]A^{\alpha_{m}}_{k'r}&\text{otherwise}\label{eq:rho--recur}.
   \end{dcases}
\end{equation}
In this relation, $\rho^{--}_{sr}(m)$ is no longer expressed in terms
of all the earlier shells as in Eq.~(\ref{eq:rho--}), but just in terms of one earlier shell, $(m-1)$. 
The recursion relation for the reduced full density matrices in
Eq.~(\ref{eq:RFDM}) may be derived in a similar way, leading to the expression
\begin{equation}
R^m_{\rm red}(k,k')=\begin{dcases}
     0 & \text{if } m=N;\\
\sum_{l\alpha_{m+1}}A^{\alpha_{m+1}}_{kl}\Big(w_{m+1}\frac{e^{-\beta E_l^{m+1}}}{Z_{m+1}}\Big)A^{\alpha_{m+1}\dagger}_{lk'}\\
+\sum_{k_1k'_1\alpha_{m+1}}A^{\alpha_{m+1}}_{kk_1}R_{\rm red}^{m+1}(k_1,k'_1)A^{\alpha_{m+1}\dagger}_{k'_1k'} & \text{otherwise}\label{eq:RFDMrecur},
   \end{dcases}
\end{equation}
in which $R^m_{\rm red}(k,k')$ depends only on terms of one later shell, $(m+1)$. 

Summarizing, one can use Eq.~(\ref{eq:rho--recur}) to determine
$\rho^{--}_{sr}(m)$ recursively in a single forward run, while
Eq.~(\ref{eq:RFDMrecur}) may be used to determine $R^m_{\rm red}(k,k')$
[and hence $\rho_{sr}^{++}(m)$ from Eq.~(\ref{eq:rho++})] recursively in a backward run. 
While the $\rho_{sr}^{++}(m)$ is the contribution to the projected
density matrix at iteration $m$ arising from lower energy states at subsequent
iterations $m'\geq m$, $\rho^{--}_{sr}(m)$ is the contribution to the
projected density matrix at iteration $m$ arising from higher energy
states at earlier iterations $m'<m$ [see Eq.~(\ref{eq:rho--})].  The
former is finite in the equilibrium limit $H^f = H^i$, yielding the
correct Boltzmann factors in thermodynamic averages, while the
latter is finite only for non-equilibrium $H^f \neq H^i$ [as can be
seen from Eq.~(\ref{eq:rho--})]. Finally, the modifications to the
recursive  expressions Eq.~(\ref{eq:rho--recur}) and Eq.~(\ref{eq:RFDMrecur})
in the case where SU(2) symmetry is used are given in Appendix~\ref{sec:su2}.
\section{Numerical Results}
\label{sec:section-4}
In this section, we apply the finite-temperature TDNRG approach to the Anderson impurity model, a prototype model of strong electronic correlations. 
Section~\ref{subsec:model} describes the model and the switching protocols that we shall consider.
Section~\ref{subsec:trace-conservation} tests the trace conservation property 
[Eq.~(\ref{eq:identity})] of the projected density matrix $\rho^{i\to f}$. This is both a useful test of the correctness of our expression for   $\rho^{i\to f}$ as
calculated from the recursion relations in Eqs.~(\ref{eq:totalRho})-(\ref{eq:RFDMrecur}), and also indicates the relative importance of the various contributions 
to this quantity at different temperatures. The time-dependence of observables is presented in Sec.~\ref{subsec:time-dependence}, where we attempt to
identify which contributions result in noise in the long-time dynamics and how this noise is affected by averaging over discretizations of the band. 
Sections~\ref{subsec:short-time-limit}-\ref{subsec:long-time-limit} present results for the short ($t\to 0^+$) and long-time ($t\to +\infty$) limits of local 
observables, such as the impurity occupation and double occupation, discussing in particular how different switching protocols, the order of switchings, 
and quench strengths affect the accuracy of the long-time limit. A comparison between results within the present TDNRG approach and that of Anders and 
Schiller in Refs.~\onlinecite{Anders2005,Anders2006}  is also presented [Fig.~\ref{fig:longshort} and Fig.~\ref{fig:short}].

\subsection{Anderson Impurity Model}
\label{subsec:model}
We illustrate the finite-temperature TDNRG method by applying it to the Anderson impurity model. This is defined by Eq.~(\ref{eq:QIS}) with
\begin{align}
&H_{\rm imp}=\sum_{\sigma}\varepsilon_{d}(t)n_{d\sigma}+U(t)n_{d\uparrow}n_{d\downarrow},\\
& H_{\rm bath}=\sum_{k\sigma}\epsilon_{k}c^{\dagger}_{k\sigma}c_{k\sigma},\\
& H_{\rm int}=\sum_{k\sigma} V(t)(c^{\dagger}_{k\sigma}d_{\sigma}+d^{\dagger}_{\sigma}c_{k\sigma}),
\end{align}
where $n_{d\sigma}$ is the number operator for electrons with spin $\sigma$ in a local level with energy $\varepsilon_{d}(t)$,  $U(t)$ is the Coulomb repulsion between electrons in this level, 
$V(t)$ is the hybridization matrix element of the local $d$ state with the conduction states, and 
$\epsilon_{k}$ is the kinetic energy of the conduction electrons with wavenumber $k$. For simplicity, we shall consider only quenches in which the $d$-level energy or Coulomb repulsion
(or both) are changed at $t=0$ while $V(t)=V$ is kept constant. Specifically, we consider, (a), switching the $d$-orbital energy, $\varepsilon_d=\theta(-t)\varepsilon_i+\theta(t)\varepsilon_f$ while keeping $U(t)=U_i=U_f$ constant, and, (b), switching the Coulomb interaction $U(t)=\theta(-t)U_i +\theta(t) U_f$ with simultaneous change of $\varepsilon_{d}(t) = \theta(-t)\varepsilon_i + \theta(t)\varepsilon_f$. In the former, the interest is in switching between different regimes of the Anderson model,
e.g., from the mixed-valence regime with $\varepsilon_{i}=0$ to the Kondo regime with $\varepsilon_f = -U_f/2$, and vice versa (or between two Kondo regimes with differing level
energies), while in the latter it is of interest to investigate switching between an uncorrelated ($U_i=0$) and a correlated system ($U_f>0$), or vice versa. 

The Anderson model has several bare energy scales, such as $\varepsilon_d, U$ and the hybridization strength $\Gamma=\pi\rho V^{2}$, where $\rho=1/W$ is the density of states per spin for a flat band of width $W=2D=2$ where $D=1$ is the half-bandwidth . In the case of strong correlations $U\gg \pi\Gamma$ and $-\varepsilon_d\gg \Gamma$ it also develops a
renormalized low energy scale, the Kondo scale $T_{\rm K}=\sqrt{U\Gamma/2}e^{\pi \varepsilon_d (\varepsilon_d +U)/2\Gamma U}$. 
Since $\varepsilon_{d}$ will often be varied below, whenever $T_{\rm K}$ appears in the numerical calculations below,
it will denote the symmetric Kondo scale with $U=\text{max}(U_{i},U_{f})$.
We shall mainly focus on the expectation value of the local level occupation 
$\langle n_{d}(t)\rangle$ and of the double occupancy $\langle n_{d\uparrow}(t)n_{d\downarrow}(t)\rangle$.
In the results reported below,  we shall express temperature and time dependences  in terms of 
$T/T_{\rm K}$ and $t\Gamma$, respectively. Short, intermediate and long times will correspond to
$t\Gamma\ll 1$, $t\Gamma\sim 1$ and $t\Gamma\gg 1$, respectively. The short-time ($t\to 0^+$) and long-time ($t\to +\infty$) limits of the tdNRG will be tested against the exactly known results 
(obtained, for example, from thermodynamic calculations for the initial/final state Hamiltonians). This allows the accuracy of the TDNRG approach to be benchmarked at finite temperature.

\begin{figure}[ht]
  \centering
    \includegraphics[width=0.51\textwidth]{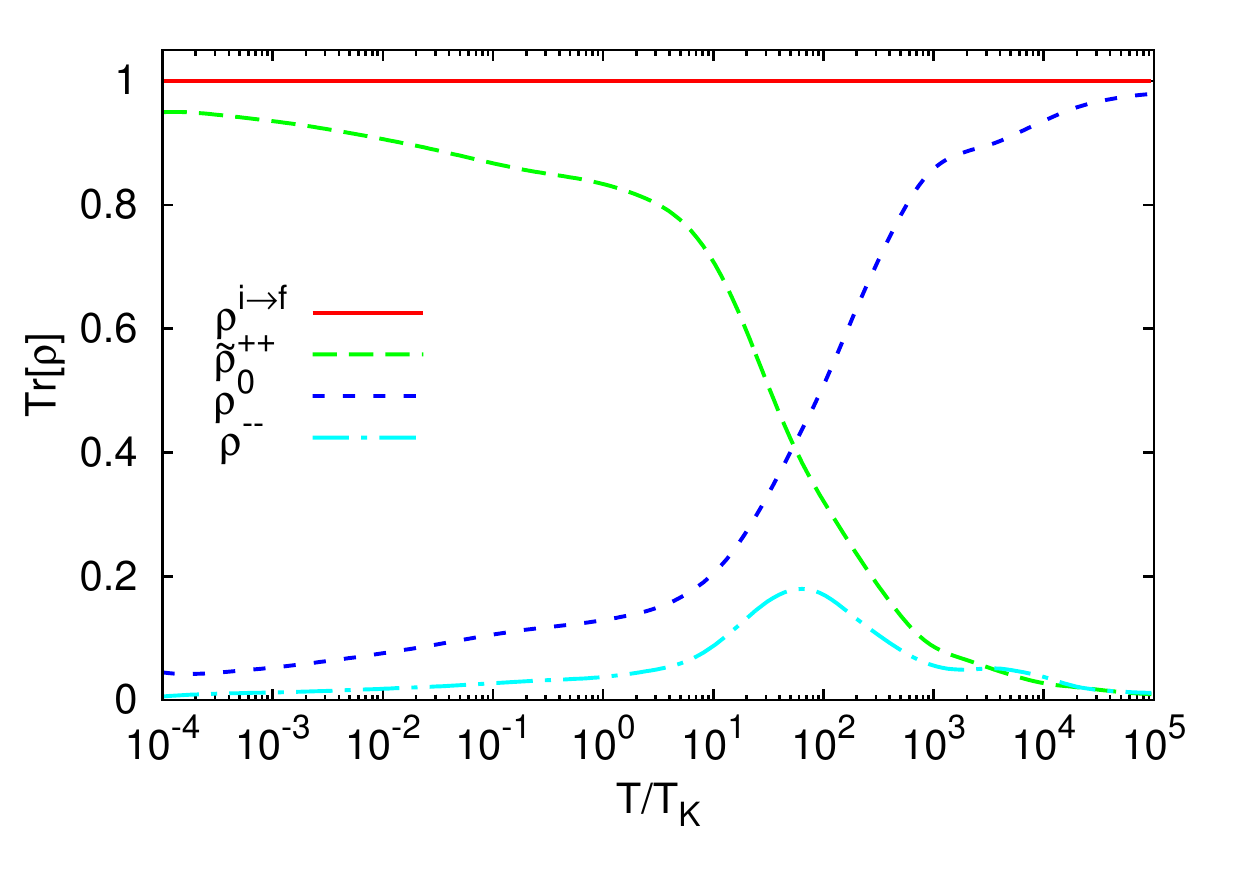}
  \caption
{
  (Color online)
  ${\rm Tr}[\rho^{i\to f}]$ and its contributions vs temperature. 
  The system is switched from $\varepsilon_i=0$ (mixed-valence regime) to $\varepsilon_f=-6\Gamma$ (symmetric Kondo regime). 
  The other parameters are $U_i=U_f=U=12\Gamma$, and $\Gamma=10^{-3}D$. $T_{\rm K}\approx 2.0\times 10^{-5}D$ is the Kondo temperature in the final state. 
  The calculations are for $\Lambda=2$, no $z$ averaging, and keeping $660$ states per NRG iteration.
}
\label{fig:traceRho}
\end{figure}
\subsection{Trace of $\rho_{sr}^{i\to f}$}
\label{subsec:trace-conservation}
Since the projected density matrix is the most important quantity in our formulation of the finite-temperature TDNRG, we consider ${\rm Tr}[\rho^{i\to f}]$ vs temperature first.  Appendix~\ref{sec:trace-formula} shows that ${\rm Tr}[\rho^{i\to f}]=1$ is an exact result, obtained for the special case of time evolution in Eq.~(\ref{eq:localOt}) when $\hat{O}=1$. 
This result is useful as a test of the numerical accuracy of the calculations as well as a way of estimating the relative contributions of 
$\tilde{\rho}^{++}$, $\rho_{0}$ and $\rho^{--}$ to $\rho^{i\to f}$ and hence their importance for the time-dependence.
Figure~\ref{fig:traceRho} shows that the trace is conserved to within an error less than $10^{-10}$ at all temperatures. 
The same accuracy holds also for any choice of $\varepsilon_d$, $U$, and $\Lambda$. The trace of each contribution to $\rho^{i\to f}$, i.e., $\tilde{\rho}^{++}$, $\rho^{0}$, and $\rho^{--}$, is also shown in Fig.~(\ref{fig:traceRho}). 
We see that ${\rm Tr}[\tilde{\rho}^{++}]$ has its main contribution at low to intermediate temperatures, while the main contribution of ${\rm Tr}[\rho^0]$ is at intermediate to high temperatures. The contribution of ${\rm Tr}[\rho^{--}]$ can be as large as $20\%$ at intermediate temperatures, but vanishes only at $T\ll T_{\rm K}$ and $T\gg T_{\rm K}$. These results exhibit the numerical precision of calculation, and also show the correctness of generalizing the TDNRG at finite temperature, which essentially depends on $\rho^{--}$.
\subsection{Time-dependence}
\label{subsec:time-dependence}
\begin{figure}[ht]
  \centering
   \includegraphics[width=0.51\textwidth]{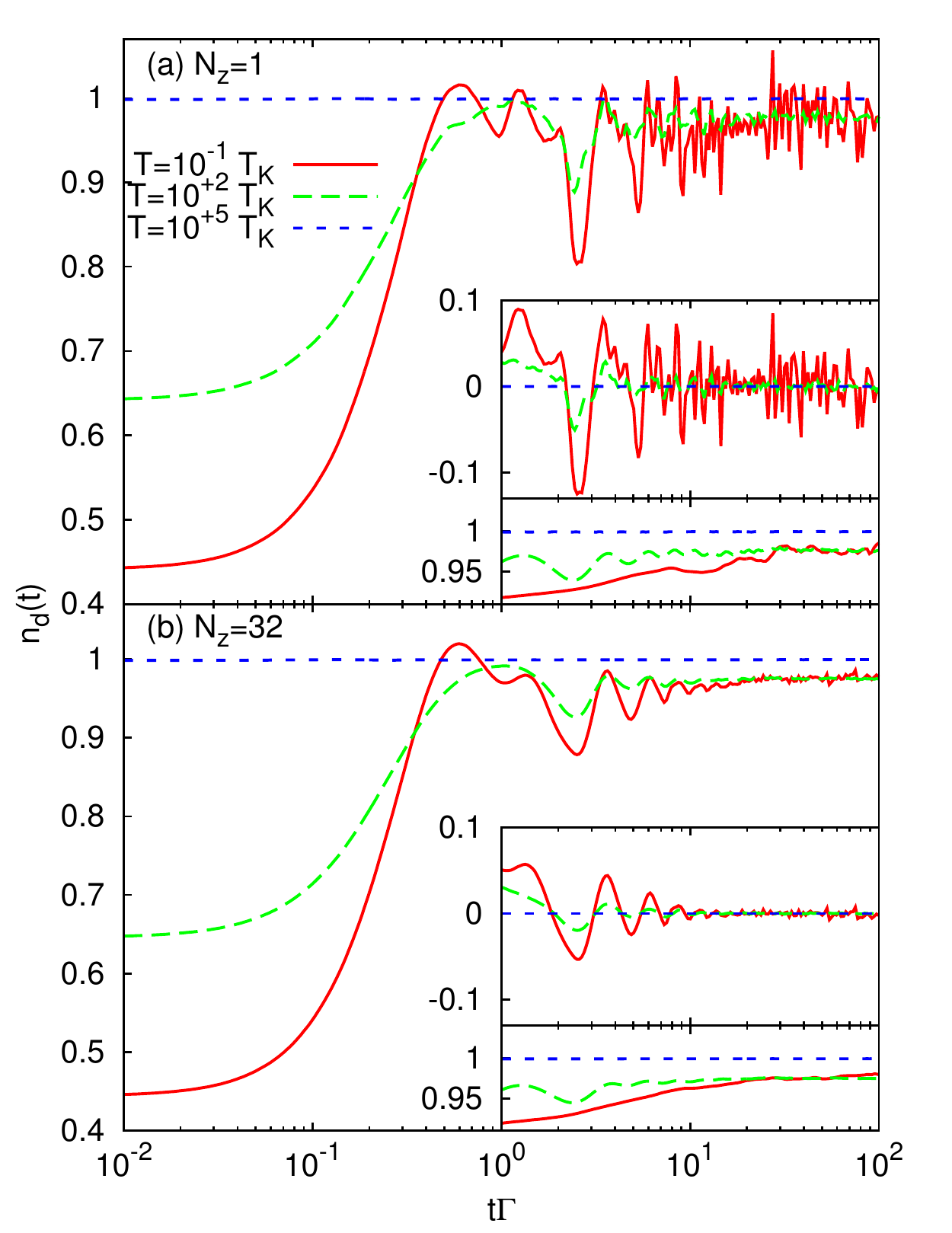}
   \caption
{
  (Color online)
  Time dependence of occupation number $n_d(t)$ at different temperatures with switching as in Fig.~\ref{fig:traceRho} 
  (i.e. from the mixed-valence to the symmetric Kondo regime with $U=U_i=U_f=12\Gamma$). 
  (a) Calculation  for $\Lambda=2$, no $z$ averaging, and keeping $660$ states per NRG iteration. 
  The upper and lower insets, respectively, represent the contributions of the kept states and discarded states to the occupation number in the range 
  $1\leq t\Gamma \leq 10^2$. 
  (b) Calculation as in (a), but with $z$ averaging, using $N_z=32$. 
  Insets as in (a) show contributions from kept (upper) and discarded (lower) states.
}
\label{fig:OvsTime}
\end{figure}

Figure~\ref{fig:OvsTime}~(a) shows the time dependence of the occupation number $n_d(t)$ after switching the system from the mixed-valence regime 
to the symmetric Kondo regime at three representative temperatures $T=0.1T_{\rm K}$, $T=100T_{\rm K}\approx 2\Gamma$ and $T=10^{5}T_{\rm K}\approx W$.
The results at lower temperatures, $T\ll T_{\rm K}$, are qualitatively similar to those in the lowest-temperature case ($T=0.1T_{\rm K}$) shown in Fig.~\ref{fig:OvsTime}, namely,
starting from the thermodynamic value of the initial state at $t=0$, $n_d(t)$ varies smoothly at short ($t\Gamma\ll 1$) and intermediate $t\Gamma\sim 1$ time scales, and develops  noise at longer times $t\Gamma \gtrsim 1$ . At higher temperature $T=100T_{\rm K} \sim 2\Gamma$ , $n_d(t)$ evolves from a higher initial value at $t=0$, as expected from the behavior of
the occupation number with temperature in the mixed-valence regime,\cite{Costi1994,Costi2010} and exhibits less noise at long times than the low temperature curve. 
Since the final state is particle-hole symmetric, one expects that $n_d(t\to +\infty)=1$ in the final state.
Instead, we see that, for the above temperatures, the long-time limit of $n_d(t)$ deviates from the exact value by $2\%-3\%$.
This will be analyzed in more detail in Sec.~\ref{subsec:long-time-limit}. Finally, we see that at  very high temperature $T=10^{5}T_{\rm K} \approx W$, the occupation number
$n_d(t)$ is almost independent of time and and lies very close to its expected thermodynamic value. The noise is also seen to be negligible in this limit. 
The main panel of Figure~\ref{fig:OvsTime}~(b) shows the effect of $z$ averaging \cite{Oliveira1994,Campo2005} with $N_z=32$ on the above results.
We see that $z$ averaging reduces the noise at long times, but does not improve on the expected long-time limit for $n_d$ (i.e., the error for the low temperature
results remains $2\%-3\%$). 

Since $\displaystyle\sum_{rs\notin KK'}=\sum_{rs\in KD}+\sum_{rs\in DK}+\sum_{rs\in DD}$, we decompose $O(t)$ in Eq.~(\ref{eq:localOt}) as follows $O(t)=O^{KD}(t)+O^{DK}(t)+O^{DD'}(t)$. $O^{KD}(t)+O^{DK}(t)$ represents the contribution from the kept states, while $O^{DD'}(t)$ represents the contribution from just the discarded states. These two contributions to the total time evolution without and with the $z$ averaging are respectively shown in the insets to Fig.~\ref{fig:OvsTime}~(a) and (b) in the time range $t\Gamma\geq 1$. The upper inset to Fig.~\ref{fig:OvsTime}~(a) shows the contribution of kept states, $n_d^{KD}(t)+n_d^{DK}(t)=2\times\sum_{m=m_0}^N \sum_{kl}^{\in KD}\rho^{i\to f}_{kl}(m) \cos[(E^m_k-E^m_{l})t] n^m_{lk}$. This contribution, evolving in time as $\cos[(E^m_k-E^m_{l})t]$, exhibits significant noise at long times. The features in the noise from this contribution are reflected in the total.
We do not observe such large noise in the remaining contribution $n_d^{DD'}(t)$, which evolves in time as $\cos[(E^m_l-E^m_{l'})t]$ (lower inset). The insets in Fig.~\ref{fig:OvsTime}~(b) also show the contributions of kept and discarded states to the total time evolution, calculated with $Nz=32$. The $z$ averaging also reduces the  noise appearing in the kept state contribution, but has a small effect on the discarded state contribution. On the other hand, $n_d^{KD}(t)+n_d^{DK}(t)$ equilibrates to zero in the long-time limit, whereas $n_d^{DD'}(t)$ makes up all the contribution in this limit (as expected, since the long-time limit consists of selecting states with $E_{r}^m=E_s^{m}$, implying that they are both discarded states). This implies that the noise is mainly coming from the eigenvalues of the kept states, or more generally, the NRG approximation, as the low-energy kept states at each iteration are the ones most affected by the NRG truncation procedure. On the other hand, the observed error in the long-time limit is likely not due to the NRG approximation, since this approximation generally has a smaller effect on the high energy (discarded) states which
are also the only ones contributing to this limit.
\begin{figure}[ht]
  \centering
   \includegraphics[width=0.51\textwidth]{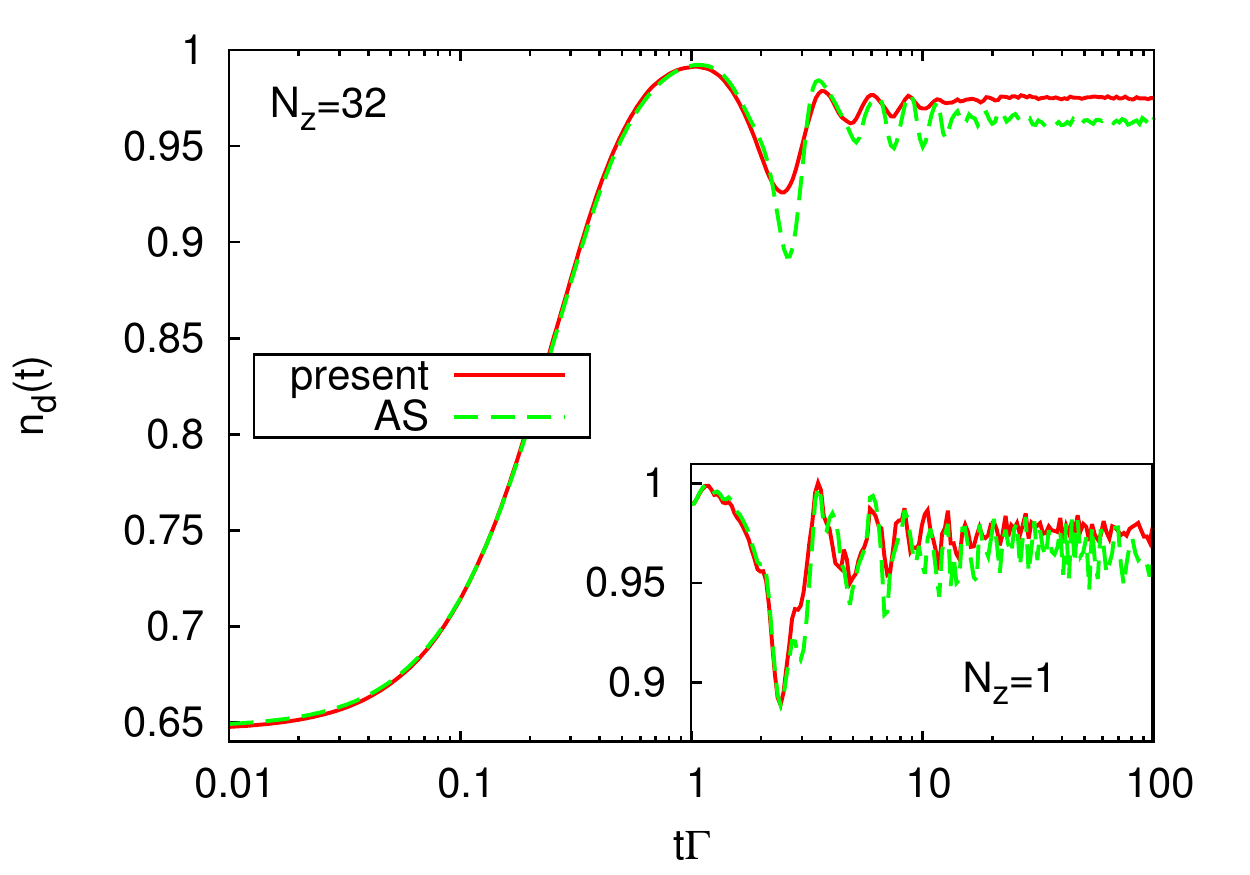}
   \caption
{
  (Color online)
  Time dependence of the occupation number $n_d(t)$ calculated within the present TDNRG (full line) and the Anders-Schiller approach (AS). 
  Switching as in Fig.~\ref{fig:traceRho} with the same parameters (mixed-valence to symmetric Kondo state). 
  The calculation is at $T=10^2T_{\rm K}$ with $\Lambda=2$ and $z$ averaging with $N_z=32$, and keeps $660$ states per NRG iteration. 
  The inset represents the results with no $z$ averaging.
}
\label{fig:longshort}
\end{figure}

It is of interest to compare the present TDNRG calculations with those from previous work\cite{Anders2005,Anders2006}. As mentioned in Sec.~\ref{sec:section-1}, previous TDNRG calculations at finite temperature $T$ were carried out by choosing a shorter chain of length $M<N$, such that $T\approx T_M=\omega_M$ where $\omega_M$ is the characteristic energy
scale of the Wilson chain of length $M$. In this approach, the initial state density matrix includes only the contribution from all the states of shell $M$ [i.e., it uses
Eq.~(\ref{eq:rho-Anders}) with $N=M$]. 
Figure~\ref{fig:longshort} shows $n_d(t)$ calculated within the present TDNRG and in the previous approach of Anders and Schiller (AS) in Refs.~\onlinecite{Anders2005,Anders2006}
at $T=10^2T_{\rm K}$, with $z$ averaging. The two calculations give very similar results at short times, while there is a clear difference at long times. 
The long-time limit of the occupation number in the present work is closer to the expected value than in the AS approach.
We also see that the time evolution in the AS approach exhibits more noise than in the present one, both with and without $z$ averaging 
(see the inset to Figure~\ref{fig:longshort}). 
These differences show that at finite temperature $T\approx T_M$, the
use of a short chain of length $M$ in the approach of AS does not capture the time evolution of local observables as accurately as within our approach based on using a FDM for the
initial state. On the other hand, at low temperatures ($T<T_{\rm K}$), the chain of length $M$ corresponding to $T$ is sufficiently long, that the two approaches give very similar results.
\subsection{Short-time limit}
\label{subsec:short-time-limit}
\begin{figure}[h]
  \centering
    \includegraphics[width=0.51\textwidth]{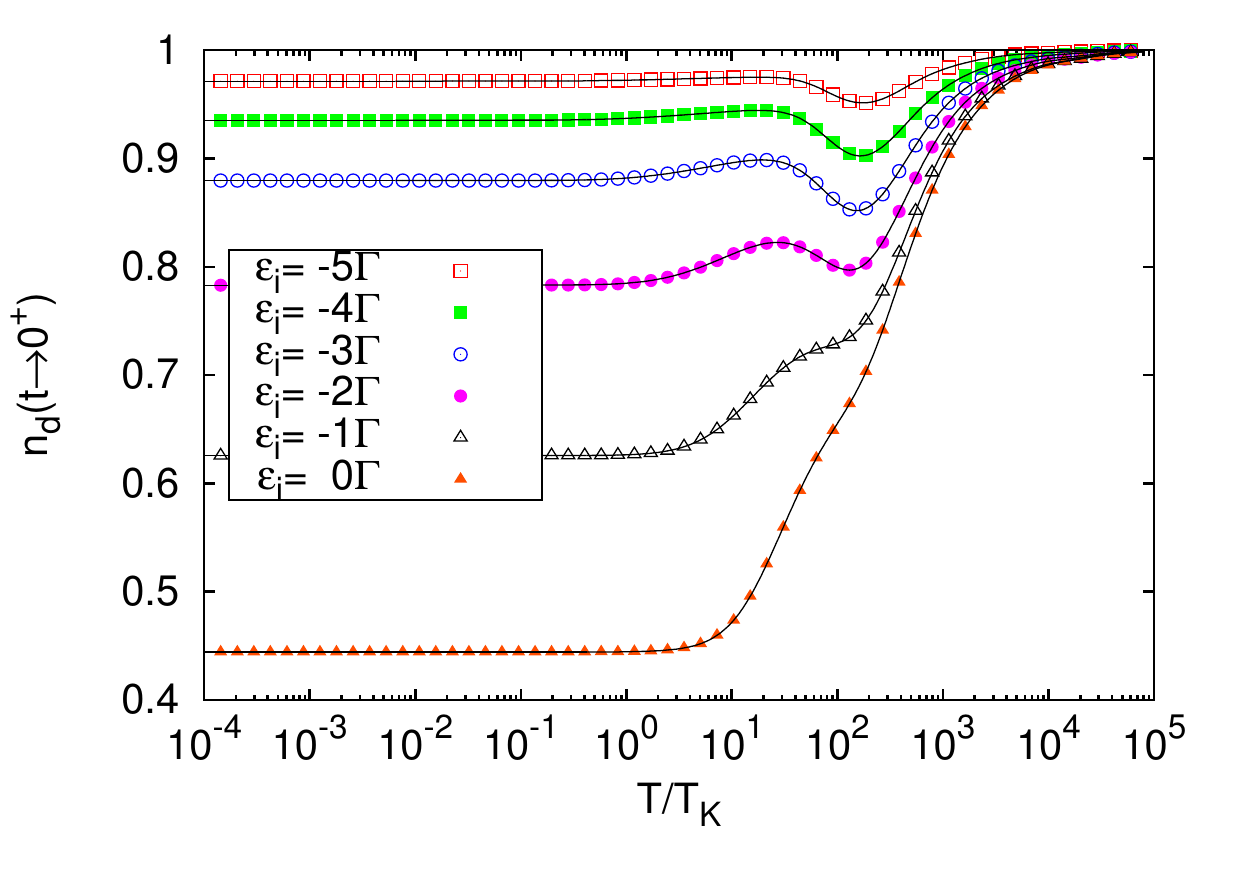}
  \caption
{
  (Color online)
  Local level occupation number $n_d(t)$ in the short-time limit vs temperature for different initial $\varepsilon_i$ keeping the final $\varepsilon_f=-6\Gamma$ and 
  $U_i=U_f=U=12\Gamma$ fixed (i.e., switching from the asymmetric model to the symmetric Kondo regime).  $\Gamma=10^{-3}D$, and $T_{\rm K}\approx 2.0\times 10^{-5}D$ 
  is the symmetric Kondo temperature of the final state. The symbols represent results with the TDNRG; the lines are the thermodynamic values in the initial state. 
  The calculations are for $\Lambda=2$, no $z$ averaging, and keeping $660$ states per NRG iteration.
}
\label{fig:Ot0}
\end{figure}

In the short-time limit, $t\to 0^{+}$, the NRG approximation is not operative in Eq.~\ref{eq:localOt}, and, as indicated in Sec~\ref{sec:section-1}, $O(t=0)$ exactly equals $O_i$, the thermodynamic value in the initial state, calculated within the FDM approach\cite{Costi2010,Merker2012b}. In Fig.~\ref{fig:Ot0}, we show the short-time limit of the occupation number 
as well as the initial state thermodynamic value in a wide range of temperatures. The system is switched from the asymmetric to the symmetric Kondo model. 
We can see that $n_d(t=0)$ within the TDNRG agrees perfectly  with the thermodynamic value at all temperatures and for any initial-state preparation, the absolute error being less than $10^{-10}$. 
We have the same precision with other sets of parameters, i.e., $\varepsilon_d$, $U$, and $\Lambda$, or other local observables, e.g., double occupancy. This provides a check on our generalization to arbitrary temperatures, on the expression for $\rho^{i\to f}$ entering the calculation of $O(t\to 0^{+})$ and on the accuracy of the numerical calculations.

We have also tested the short-time limit within the AS approach for the same system and switching parameters as in Fig.~\ref{fig:Ot0}. In this case,
the short-time limit for $n_{d}(t)$ corresponds exactly to the thermodynamic value in the initial state, calculated within the conventional approach to thermodynamics
via the NRG \cite{Oliveira1981,Merker2012a}. As shown in Ref.~\onlinecite{Merker2012b}, the FDM and conventional approaches to thermodynamics 
give essentially the same results, so we conclude that the short-time limit of the AS approach is also numerically exact.

\subsection{Long-time limit and thermalization}
\label{subsec:long-time-limit}
\begin{figure}[h]
  \centering
    \includegraphics[width=0.51\textwidth]{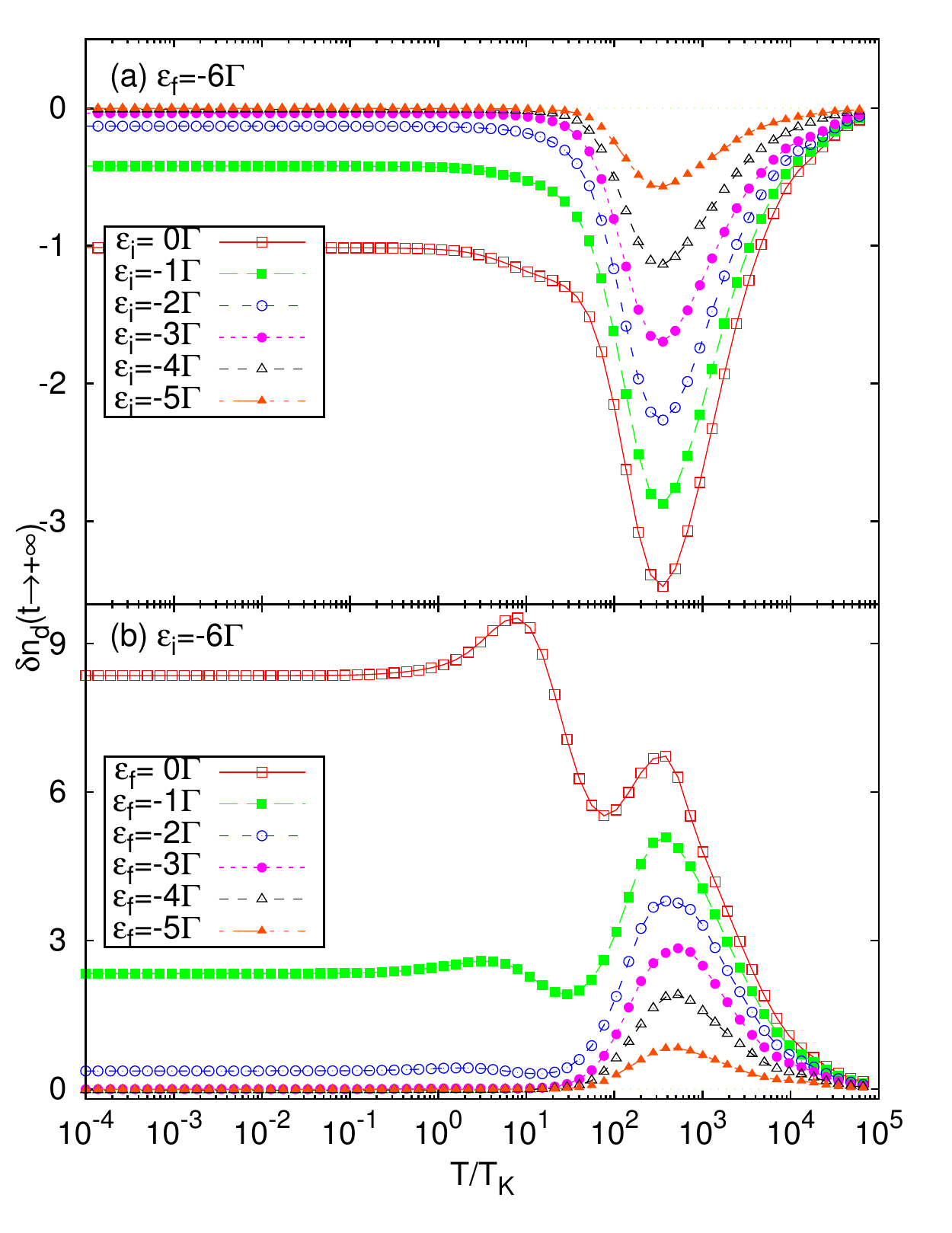}
  \caption
{
  (Color online)
  The (percentage) relative error of the local level occupation number in the long-time limit vs temperatures. $U_i=U_f=U=12\Gamma$ and $\Gamma=10^{-3}D$. 
  (a) Switching is from the asymmetric model to the symmetric Kondo regime, as in Fig.~\ref{fig:Ot0}.
  (b) Reverse switching, from the symmetric Kondo regime to the asymmetric model [the $\varepsilon_f$ are indicated and are the same as the $\varepsilon_i$ in (a)]. 
  $T_{\rm K}\approx 2.0\times 10^{-5}D$ is the Kondo temperature of the symmetric model
  and calculations are for $\Lambda=1.6$, no $z$ averaging, and keeping $660$ states per NRG iteration.
}
\label{fig:Otinf}
\end{figure}

In the long-time limit, $t\to +\infty$, we already clarified in the previous section that the NRG approximation is not the main source of error . However, there is another source of error, the logarithmic discretization of the band. In order to estimate the accuracy of the long-time limit, we define the following percentage relative error
\begin{equation}
\delta O(t\to +\infty)=100\times \frac{O(t\to +\infty)-O_f}{O_f}\label{eq:percenterror},
\end{equation}
in which $O_f$ is the thermodynamic value in the final state calculated within the FDM approach to thermodynamics.\cite{Merker2012b} The closeness of the long-time limit to $O_f$ indicates either the extent to which
the system thermalizes at long times, or the extent of the errors arising from the discretization (the two may also be related \cite{Rosch2012}).

Figure~\ref{fig:Otinf} shows the relative error in $n_{d}(t\to +\infty)$ upon switching from the asymmetric to the symmetric model [Fig.~\ref{fig:Otinf}~(a)], 
and vice versa [Fig.~\ref{fig:Otinf}~(b)]. We can see two trends. The first one is that the larger the quench, measured by the difference $\Delta \varepsilon_d$
between $\varepsilon_i$ and $\varepsilon_f$ in this case, the larger the error in the long-time limit in both Figs.~\ref{fig:Otinf}~(a) and \ref{fig:Otinf}~(b). The
error in the former lies below $4\%$ in all cases and in the latter below $10\%$. The error vanishes for $\Delta\varepsilon_d\to 0 $ and one recovers the 
exact equilibrium results. The error  exhibits an extremum at a temperature $T\approx 300T_{\rm K}\approx 6\Gamma=U/2$ in both Figs.~\ref{fig:Otinf}~(a) and \ref{fig:Otinf}~(b).
In addition, a second extremum is visible for large quenches in  Fig.~\ref{fig:Otinf}~(b) at $T\approx 7T_{\rm K}=0.14\Gamma$. Signatures of this, as shoulders, are present also in Fig.~\ref{fig:Otinf}~(a) for the largest quenches.
The second trend is the strong dependence of the error on the magnitude of the largest incoherent excitation in the final state,
denoted by $\varepsilon_{\rm inc}^{\rm max}$ and defined by
\begin{align}
& \varepsilon_{\rm inc}^{\rm max} = \text{max}( |\varepsilon_f|,|\varepsilon_f+U|,\Gamma).\label{eq:incoherent}
\end{align}
We see that $\varepsilon_{\rm inc}^{\rm max}=6\Gamma$ for all final states of  Fig.~\ref{fig:Otinf}~(a), but increases from $7\Gamma$ to $12\Gamma$ for the
corresponding final states of Fig.~\ref{fig:Otinf}~(b), with the error increasing monotonically with increasing $\varepsilon_{\rm inc}^{\rm max}$.

\begin{figure}[h]
\centering
\includegraphics[width=0.51\textwidth]{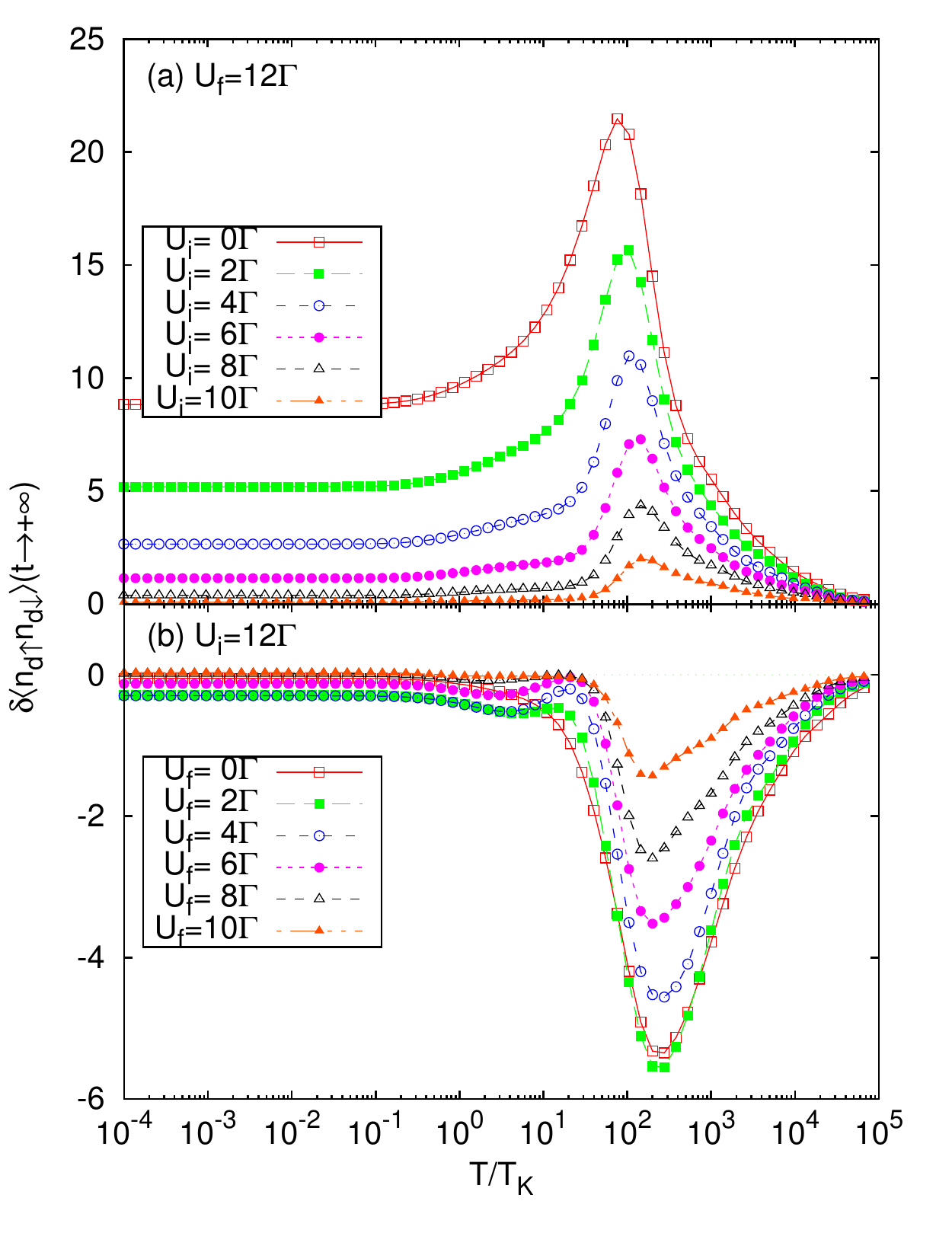}
\caption
{
  (Color online)
  The (percentage) relative error of the double occupation number in the long-time limit vs temperature. 
  $\Gamma=10^{-3}D$, and switching is applied maintaining the particle-hole symmetry, i.e., $\varepsilon_d(t)+U(t)/2=0$. 
  (a) Switching from a less correlated system to a more correlated one ($U_f>U_i$). 
  (b) Reverse switching, from a more correlated to a less correlated system ($U_f<U_i$). 
  The $U_f$ are indicated and are the same as the $U_i$ in (a). 
  $T_{\rm K}\approx 2.0\times 10^{-5}D$ is the Kondo temperature of the most correlated system and calculations are for $\Lambda=1.6$, 
  no $z$ averaging, and keeping $660$ states per NRG iteration.
}
\label{fig:figure6}
\end{figure}

We consider a different switching protocol in order to check that the above two trends persist. We switch both the Coulomb interaction $U(t)=\theta(-t)U_i+\theta(t)U_f$, and the level energy $\varepsilon_d(t)$ such that $\varepsilon_d(t)+U(t)/2=0$, thereby maintaining particle-hole symmetry in both the initial and final states. By doing so, we are considering switching from a strongly correlated system to a less correlated one (or vice versa). Since $\varepsilon_d(t)+U(t)/2=0$, the occupation number equals $1$ in both initial and final states, and we find numerically that 
$n_d(t)=1$ with negligible error. Therefore, we present here the results of the double occupancy in the long-time limit. 
Figure~\ref{fig:figure6}~(a) shows the relative error in the long-time limit upon switching from a less correlated to a more correlated system and Fig.~\ref{fig:figure6}~(b) shows the same error but with the switching reversed.  We see that both cases follow the trend that the larger the quench, $\Delta U =|U_f-U_i|$, the bigger the error, with a small violation of this trend only
for the case of switching to an uncorrelated system with $U_f=0$ in Fig.~\ref{fig:figure6}~(b) where the error is almost the same as for the next most correlated system with $U_f=2\Gamma$. We also note that the extrema in the error for the double occupancy are qualitatively similar to those 
found for the occupation number in Figs.~\ref{fig:Otinf}~(a) and \ref{fig:Otinf}(b) above. Comparing now the same size of quench, in both Figs.~\ref{fig:figure6}~(a) and \ref{fig:Otinf}(b), we see that the error 
is larger in Fig.~\ref{fig:figure6}~(a) with $\varepsilon_{\rm inc}^{\rm max}$ constant at $6\Gamma$, than in  Fig.~\ref{fig:figure6}~(b) with  $\varepsilon_{\rm inc}^{\rm max}$ smaller and ranging from $\Gamma$
to $5\Gamma$. We note also that the errors in the long-time limit of the double occupancy are larger than the corresponding ones in the occupation discussed above, reaching values as
large as $25\%$ for highly correlated final states. This is due to the fact that the double occupancy can become very small in a correlated system, hence while the absolute errors in the
long-time limit for different observables are actually found to be comparable, the relative error for the double occupancy can become large for highly correlated final states.

Combining the observations from Fig.~\ref{fig:figure6} with those of Fig.~\ref{fig:Otinf}, we conclude that the error in the long-time limit of a local observable depends strongly
on, (a), the size of the quantum quench, and, (b), the magnitude of the highest energy incoherent excitation in the final state. A possible reason for (b), is that
the logarithmic discretization scheme used in the TDNRG calculations does not capture accurately enough the high energy excitations (see the next paragraph for supporting results). 
Further discussion of the long-time limit, supporting the above conclusions, is presented in Appendix~\ref{sec:supporting},  where we show results for 
additional switching protocols.

\begin{figure}[h]
  \centering
    \includegraphics[width=0.51\textwidth]{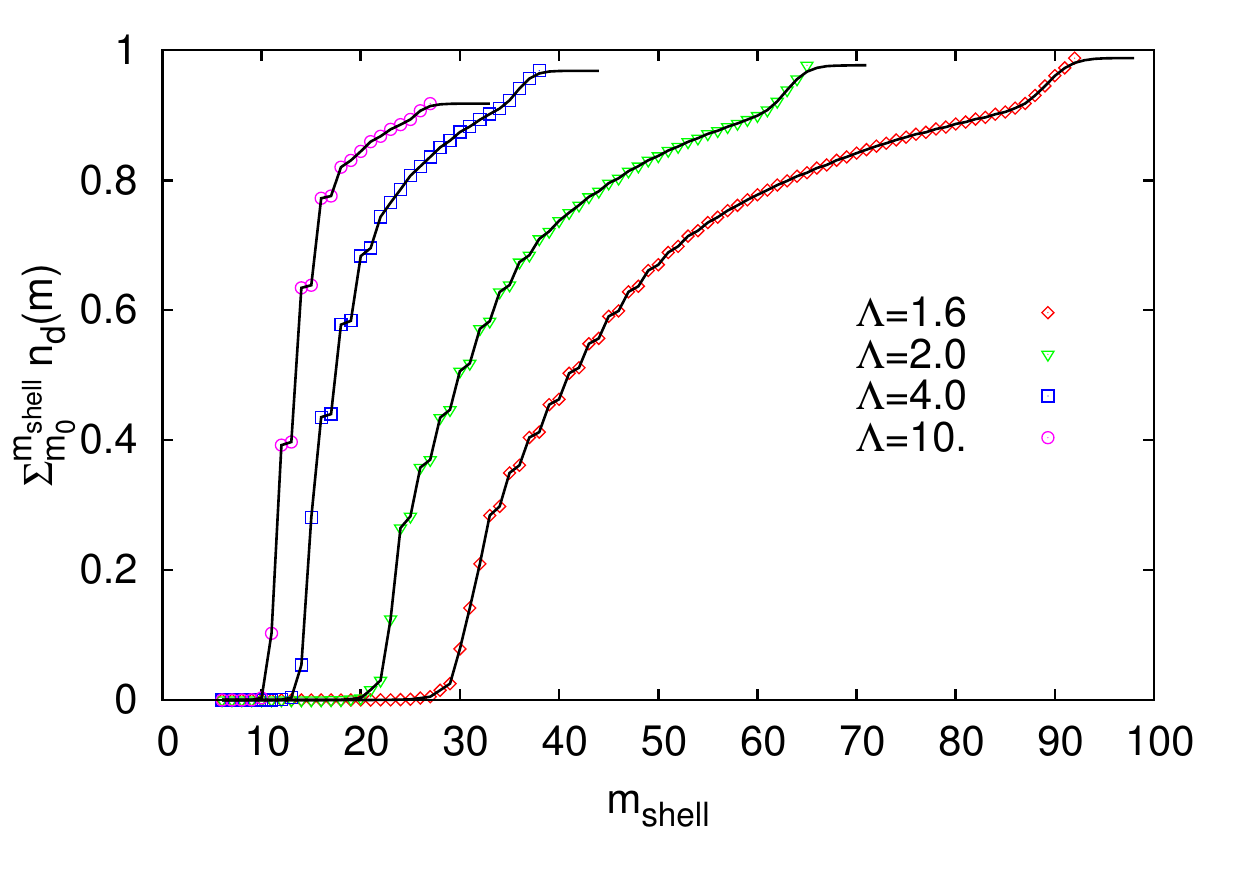}
  \caption
{
  (Color online)
  $\Lambda$ dependence of the shell accumulated local level occupation number in the long-time limit at $T=10^{-4}T_{\rm K}$ (symbols). Switching is from the
  mixed-valence ($\varepsilon_i=0$) to the symmetric Kondo regime ($\varepsilon_f=-6\Gamma $) with $U=U_i=U_f=12\Gamma$. The Kondo scale in the final
  state is $T_{\rm K}\approx 2.0\times 10^{-5}D$. The calculations are done with no $z$ averaging and keeping $660$ states per NRG iteration. The solid lines indicate the effect of increasing the Wilson chains above by six additional sites, maintaining
the same $T$. The total $n_{d}(t\to\infty)=\sum_{m=m_{0}}^{N'=N+6}n_d(m)$ is unchanged, indicating converged results (see the text
for further discussion).
}
\label{fig:total}
\end{figure}

In order to support the point made above concerning the logarithmic discretization, we consider observables in the long-time limit, calculated with different discretization parameters, $\Lambda$, and at $T=10^{-4}T_{\rm K}$ (essentially zero temperature). The system considered is that in Fig.~\ref{fig:OvsTime}~(a), with switching being from the mixed-valence to the symmetric Kondo regime. The expected value of the occupation number in the long-time limit is $1$. Fig.~\ref{fig:total} shows the shell accumulated value for $n_d(t\to +\infty)$, i.e. the quantity $\sum_{m=m_0}^{m_{\text{shell}}}n_d(m)$ where $n_d(m)$ is the contribution to the long-time limit from shell $m$. We observe that the final result for $m_{\text{shell}}=N$, with contributions from all shells included,  $n_d(t\to +\infty)$, is closer to the expected value of $1$ when $\Lambda$ is closer to $1$, the continuum limit. A smaller $\Lambda$ resolves the spectrum better at high 
energies and could account for the observation (b) above.  The trend with decreasing $\Lambda$ suggests that the exact result can be obtained in the limit $\Lambda\to 1^+$. 
\begin{figure}[h]
  \centering
    \includegraphics[width=0.51\textwidth]{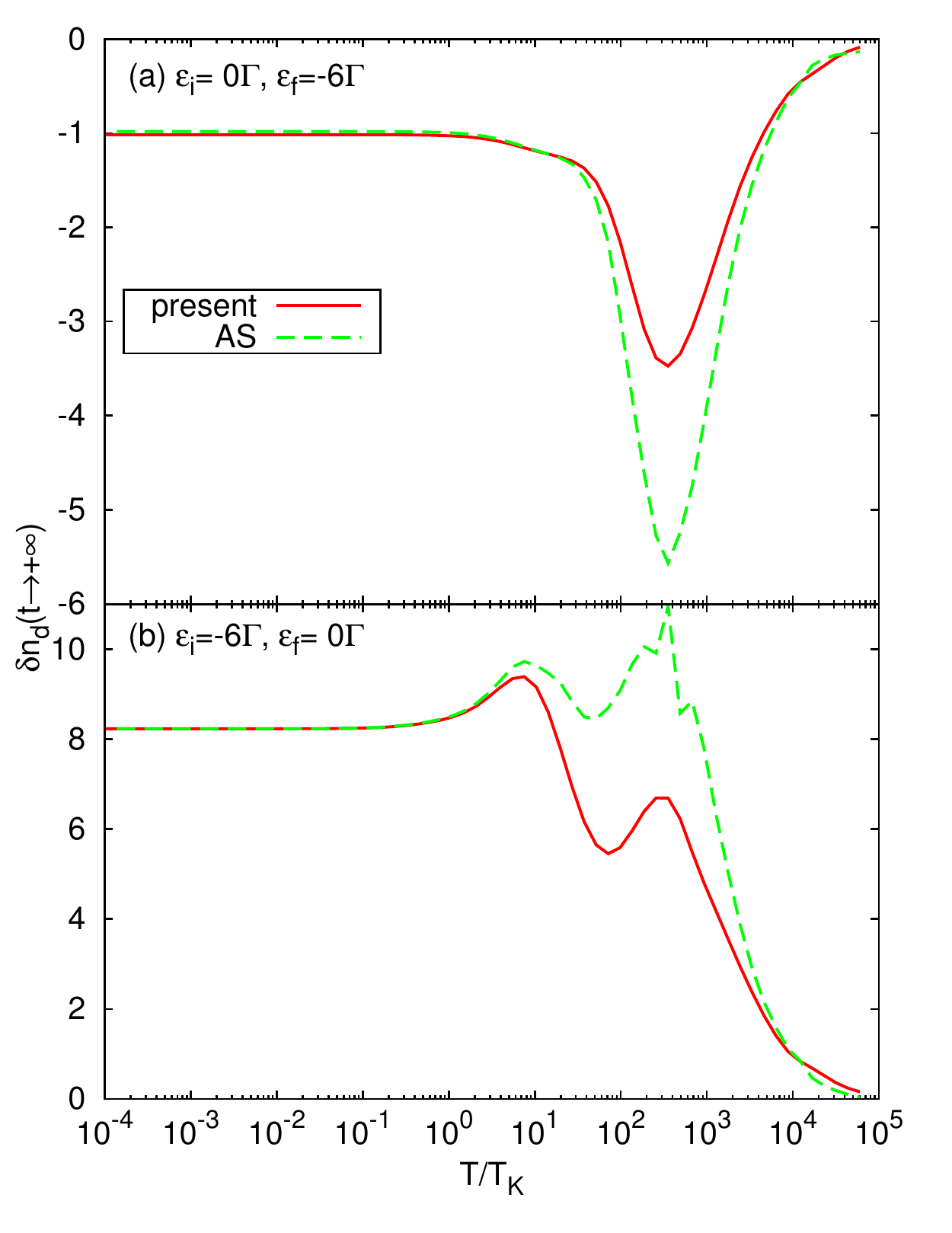}
  \caption
{
  (Color online)
  The (percentage) relative error of the local level occupation number in the long-time limit calculated with the present TDNRG (full line) and the approach of Ref.~\onlinecite{Anders2005,Anders2006} (AS) at all temperatures. 
  (a) The system is switched from the mixed-valence regime with $\varepsilon_i= 0$ to the symmetric Kondo regime with $\varepsilon_f=-6\Gamma$ keeping $U_i=U_f=U=12\Gamma$ fixed and $\Gamma=10^{-3}D$. The calculation is for $\Lambda=1.6$ and $N_z=32$, and keeps $660$ states per NRG iteration. 
  (b) As in (a), but with the reverse switching from the symmetric Kondo ($\varepsilon_i=-6\Gamma$) to the mixed-valence ($\varepsilon_f= 0\Gamma$) regime. 
  Other parameters are as in (a) and $T_{\rm K}\approx 2.0\times 10^{-5}D$ denotes the symmetric Kondo scale. 
}
\label{fig:short}
\end{figure}

It is worth commenting on whether the results for the total occupation number $n_{d}(t\rightarrow\infty)=\sum_{m=m_{0}}^{N}n_d(m)$ shown in Fig.~\ref{fig:total} (symbols), are
indeed converged with respect to the chain length $N=N(\Lambda)$ used in the calculations.  For each $\Lambda$
we choose $N$ such that $\Lambda^{-(N-1)/2}\approx 10^{-4}T_{\rm K}$ which is sufficiently long to
carry out calculations for all temperatures down to the smallest scale accessible, i.e., down to $T\approx 10^{-4}T_{\rm K}$, the temperature used in Fig.~\ref{fig:total}. Nevertheless, in order to show that the above results 
are indeed converged, and that the observed trend that $n_{d}(t\rightarrow\infty)$ comes closer to the exact
value with decreasing $\Lambda$ is correct, we carried out calculations also for longer Wilson chains and showed that
$n_{d}(t\rightarrow \infty)$ remained unchanged. This is illustrated by the solid lines in Fig.~\ref{fig:total}, which show the shell accumulated occupation numbers for a Wilson chain of length $N'=N+6$ with $T=10^{-4}T_{\rm K}$ fixed. The final value of $n_{d}(t\rightarrow \infty)$ at shell $N'$ is unchanged, but the approach of the shell accumulated occupation to this value is modified. The upturn in the contribution to an observable from shells close
to $m=N_T$ where  $\Lambda^{-(N_T-1)/2}\approx T$ is a feature of multiple shell calculations, 
\cite{Weichselbaum2007} but does not imply that the cumulative sum over all shells is not converged.

Finally, we compare the long-time limit of the present TDNRG with previous work in Refs.~\onlinecite{Anders2005,Anders2006}. Figure~\ref{fig:short} exhibits the relative error of the occupation number in the long-time limit with, (a), switching from the mixed-valence to the symmetric Kondo regime, and, (b), vice versa. This figure illustrates further details of the comparison discussed in Sec.~\ref{subsec:time-dependence} that at low temperatures $T<T_{\rm K}$ the  two schemes give essentially the same results, while, a significant difference arises in the temperature range $T_{\rm K}\leq T<10^4T_{\rm K}$, with the present TDNRG calculations showing improved accuracy  for both Fig.~\ref{fig:short}~(a) and (b).
\section{Multiple Quenches and General Pulses}
\label{sec:generic}
\begin{figure}[h]
\centering
\includegraphics[width=0.51\textwidth]{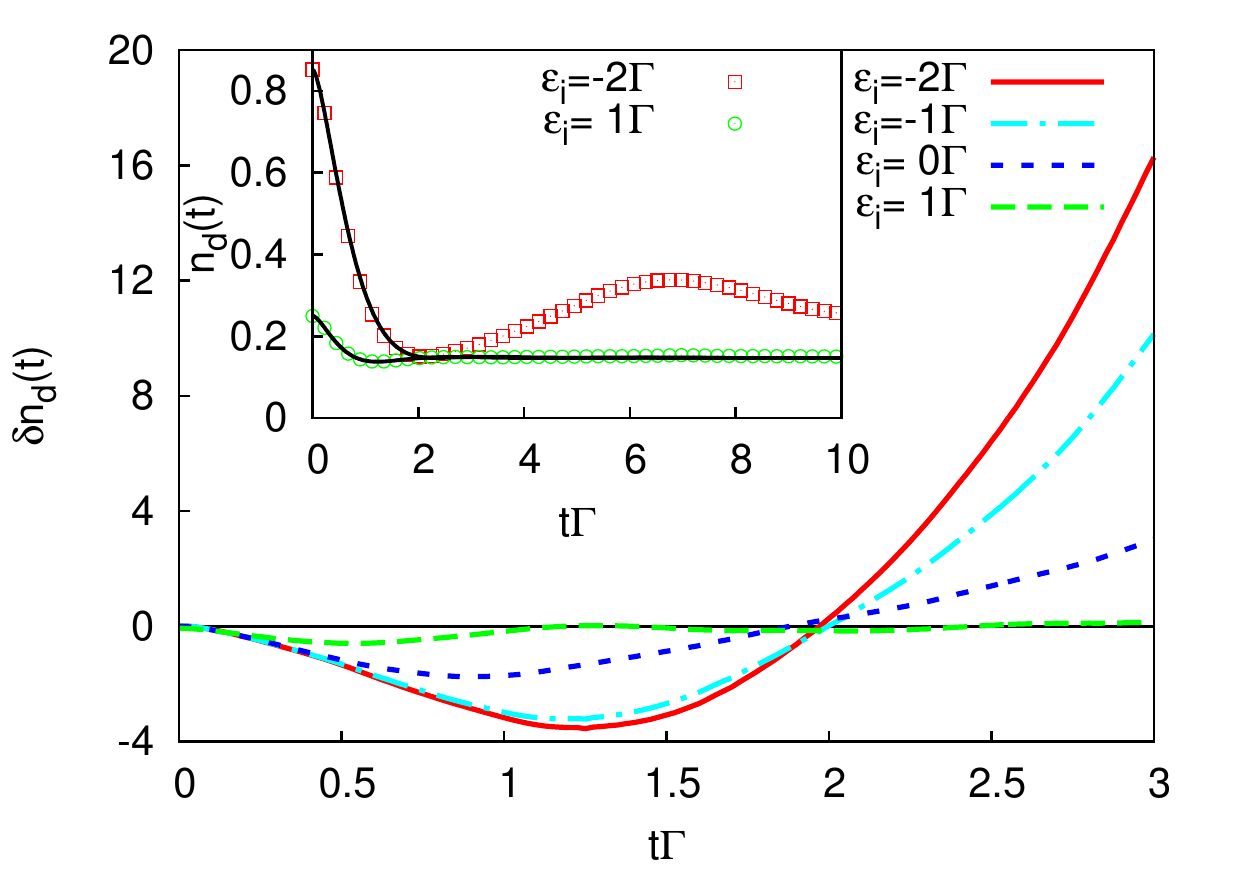}
\caption
{
  (Color online)
  Percentage relative error in the occupation number calculated with TDNRG applied to the resonant level model, using the exact analytic results for reference. 
  Calculations are with varying $\varepsilon_i$ and fixed $\varepsilon_f=2\Gamma$. $D=10^3\Gamma$, $T= 10^{-4}\Gamma$, $\Lambda=2$, and $N_z=32$. 
  The inset shows the time evolution of the occupation number calculated within TDNRG (dots) and the analytic expressions (solid lines) in the cases that the 
  quenches are the biggest and smallest in the main figure.
}
\label{fig:tdNRGvsAnalytic}
\end{figure}

From the results in Sec.~\ref{sec:section-4} we see that the TDNRG approach
becomes increasingly more accurate at long times with decreasing size of the quantum quench, while
in the short-time limit  $t\rightarrow 0^{+}$ it is always exact [see Eq.~\ref{eq:Ot0} and Fig.~\ref{fig:Ot0}]. We can also show
that the TDNRG calculation remains more accurate at short to intermediate times with decreasing quench size. To show this, we require
a comparison to exact results at finite time. These are available for the non-interacting limit of the Anderson model
$U_i=U_f=U=0$, i.e. for the resonant level model.\cite{Anders2006} In Fig.~\ref{fig:tdNRGvsAnalytic} we show the 
(percentage) relative error $\delta n_d(t)$ in $n_d(t)$ from TDNRG calculations, 
using the analytic result for the resonant level model as reference. The error is shown for several quantum quenches
corresponding to varying $\varepsilon_i$ keeping $\varepsilon_f=2\Gamma$ fixed, i.e. the quench sizes are
$\Delta\varepsilon_d=\varepsilon_f-\varepsilon_i=\Gamma,2\Gamma,3\Gamma,4\Gamma$. 
From Fig.~\ref{fig:tdNRGvsAnalytic} we see that for sufficiently small quantum quenches $\Delta\varepsilon_d\leq \Gamma$, the relative error in $n_d(t)$ remains below $1\%$ for times exceeding $t\Gamma\sim 1$, whereas for larger quenches the error exceeds $1\%$ at $t\Gamma\approx 0.5$. 
This is further illustrated in the inset to Fig.~\ref{fig:tdNRGvsAnalytic}, which shows $n_d(t)$ for the largest and smallest quenches. 
The analytic results in both quenches saturate to the same value, lying on top of each other at long times 
(since the final state energy $\varepsilon_f$ is fixed for all quenches), whereas we see that the TDNRG result in the case of the 
smaller quench fits very well the analytic one in the whole time interval, while, in the case of 
the larger quench, the TDNRG result fits the analytic one only up to $t\Gamma\lesssim 2$, and deviates from this at longer times.

\begin{figure}[ht]
\centering
\includegraphics[width=0.4\textwidth]{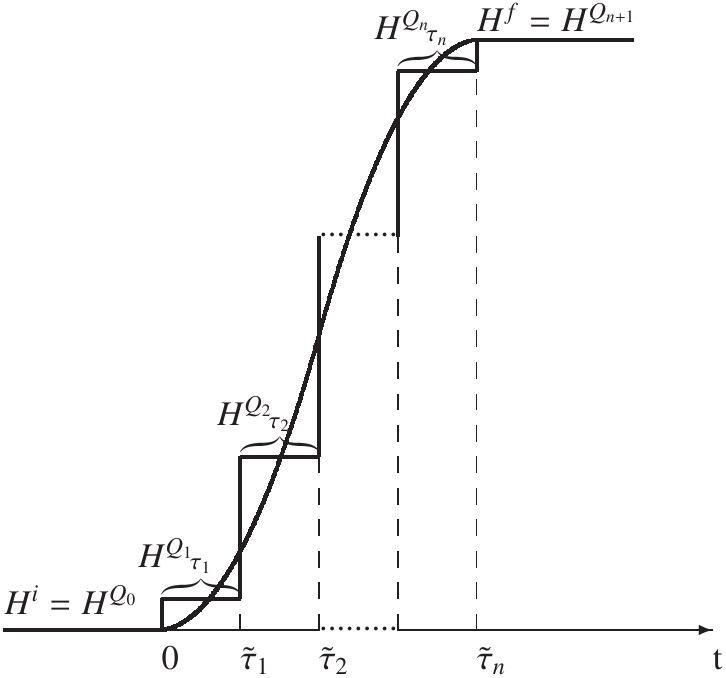}
\caption
{
  A system driven from an initial to a final state via a sequence of quantum quenches at times $\tilde{\tau}_0=0, \tilde{\tau}_1,\dots,\tilde{\tau}_n$
  with evolution according to $H^{Q_p}$ in the time interval $\tilde{\tau}_p>t\geq\tilde{\tau}_{p-1}$. 
  Such a sequence of multiple quantum quenches could also be used to describe periodic switching or to 
  approximate any general continuous pulse (smooth solid line).
}
\label{fig:generic}
\end{figure}

In summary, then, the TDNRG is a controlled method for short times and sufficiently small quantum quenches. This suggests a way to improve the long-time 
limit within a TDNRG approach, by replacing a large quantum quench by a sequence of smaller quantum quenches 
over a short time scale $\tilde{\tau}_n$. The time evolution from the initial state at $t<0$ to the final state at $t>\tilde{\tau}_n$ is achieved by applying the
TDNRG to the sequence, $i=1,\dots,n$, of small quantum quenches in the intervals $\tilde{\tau}_i\le t <
\tilde{\tau}_{i+1}$ with $\tilde{\tau}_0=0$ and $\tilde{\tau}_n=\sum_{i=1}^{n}\tau_i$ (see Fig.~\ref{fig:generic}). For sufficiently small intervals, this should then allow the TDNRG to
access longer times more accurately than would be possible with a single large quantum quench. 

In this section we show how the above idea of using multiple quenches to improve the long-time limit may be accomplished 
within the formalism presented in Sec.~\ref{sec:section-2}-\ref{sec:section-3}. 
As illustrated schematically in  Fig.~\ref{fig:generic}, such an approach would also allow the treatment of general 
continuous pulses acting in a finite time interval $\tilde{\tau}_n$, since a sequence of small quantum quenches acts as a discrete
approximation for a continuous pulse. Such pulses, acting over a finite time interval, also correspond more closely to the actual situation in 
experiments. As a special case of the general multiple quench that we shall treat, we mention the case of periodic switching between two states 
(i.e., a train of square pulses).\cite{Eidelstein2012}

For a system driven through a set of quenches, as in Fig.~\ref{fig:generic}, the time evolved density matrix at a general time in the interval $\tilde{\tau}_{p+1}>t\geq\tilde{\tau}_p$ is given by
\begin{align}
\rho(t)=e^{-iH^{Q_{p+1}}(t-\tilde{\tau}_p)}e^{-iH^{Q_p}{\tau}_p}...e^{-iH^{Q_1}{\tau}_1}\rho e^{iH^{Q_1}{\tau}_1}...e^{iH^{Q_p}{\tau}_p}e^{iH^{Q_{p+1}}(t-\tilde{\tau}_p)},
\end{align}
in which $H^{Q_{p}},p=1,\dots,n$ are intermediate quench Hamiltonians, acting during time intervals of length $\tau_{p},p=1,\dots,n$, that 
determine the time evolution at intermediate times and $H^{Q_0}=H^{i}$ and $H^{Q_{n+1}}=H^f$ are the initial and final state Hamiltonians, respectively 
(see Fig.~\ref{fig:generic}).

The time evolution of a local observable at $\tilde{\tau}_{p+1}>t\geq\tilde{\tau}_p$ is expressed in terms of the complete basis set of $H^{Q_{p+1}}$ as
\begin{align}
&O(t)=\sum_{mle} {_{Q_{p+1}}}\langle lem|\rho(t)\hat{O}| lem\rangle_{Q_{p+1}}\nonumber\\
&=\sum_{mle} {_{Q_{p+1}}}\langle lem|e^{-iH^{Q_{p+1}}(t-\tilde{\tau}_p)}...e^{-iH^{Q_1}{\tau}_1}\rho e^{iH^{Q_1}{\tau}_1}...e^{iH^{Q_{p+1}}(t-\tilde{\tau}_p)}\nonumber\\
&\hspace{16em}\times\hat{O}| lem\rangle_{Q_{p+1}},
\end{align}
in which we used $\sum_m$ for $\sum_{m=m_0}^N$. Inserting $1^{Q_{p+1}}=\sum_{mle} | lem\rangle_{Q_{p+1}}{_{Q_{p+1}}}\langle lem|$, the identity belonging to $H^{Q_{p+1}}$, in front of the operator $\hat{O}$ in the above equation, and using $\sum_{ke}| kem\rangle \langle kem|=\sum_{m'=m+1}^N \sum_{l'e'}| l'e'm'\rangle \langle l'e'm'|$, we have 
\begin{align}
O(t)=&\sum_{mrse}^{\notin KK'} {_{Q_{p+1}}}\langle rem|e^{-iH^{Q_p}{\tau}_p}...e^{-iH^{Q_1}{\tau}_1}\rho e^{iH^{Q_1}{\tau}_1}...e^{iH^{Q_p}{\tau}_p}| sem\rangle{_{Q_{p+1}}} \nonumber\\
&\hspace{12em}\times e^{-i(E^m_r-E^m_s)(t-\tilde{\tau}_p)} O^m_{sr}\nonumber\\
=&\sum_{mrs}^{\notin KK'}\rho^{i\to Q_{p+1}}_{rs}(m,\tilde{\tau}_p) e^{-i(E^m_r-E^m_s)(t-\tilde{\tau}_p)} O^m_{sr}\label{eq:Otime},
\end{align}
with 
\begin{align}
&\rho^{i\to Q_{p+1}}_{rs}(m,\tilde{\tau}_p)\nonumber\\
&= \sum_e{_{Q_{p+1}}}\langle rem|e^{-iH^{Q_p}{\tau}_p}...e^{-iH^{Q_1}{\tau}_1}\rho e^{iH^{Q_1}{\tau}_1}...e^{iH^{Q_p}{\tau}_p}| sem\rangle{_{Q_{p+1}}}.\nonumber
\end{align}
Evidently, this projected density matrix can recover its counterpart for a single quench since $\rho^{i\to Q_{1}}_{rs}(m,\tilde{\tau}_0)=\sum_e{_{Q_{1}}}\langle rem|\rho| sem\rangle{_{Q_{1}}}=\rho^{i\to Q_{1}}_{rs}(m)$. Substituting the FDM of the initial state into the above projected density matrix, we have
\begin{align}
&\rho^{i\to Q_{p+1}}_{rs}(m,\tilde{\tau}_p)=\sum_{m_1l_1e_1e}{_{Q_{p+1}}}\langle rem|e^{-iH^{Q_p}{\tau}_p}...e^{-iH^{Q_1}{\tau}_1}|l_1e_1m_1\rangle{_i}\nonumber\\
&\hspace{3em}\times w_{m_1} \frac{e^{-\beta E_{l_1}^{m_1}}}{\tilde{Z}_{m_1}}{_i}\langle l_1e_1m_1|e^{iH^{Q_1}{\tau}_1}...e^{iH^{Q_p}{\tau}_p}| sem\rangle_{Q_{p+1}}\label{eq:rhomultiquenches}.
\end{align}
We decompose $\rho^{i\to Q_{p+1}}_{rs}(m,\tilde{\tau}_p)$ into three terms,
\begin{align}
\rho^{i\to Q_{p+1}}_{rs}(m,\tilde{\tau}_p)=&\tilde{\rho}^{++}_{rs}(m,\tilde{\tau}_p)+\rho^{0}_{rs}(m,\tilde{\tau}_p)+\rho^{--}_{rs}(m,\tilde{\tau}_p),\label{eq:Totalrhomultiquenches}
\end{align}
corresponding to the $m_1>m$, $m_1=m$, and $m_1<m$ contributions in Eq.~(\ref{eq:rhomultiquenches}) [compare with Eq.~(\ref{eq:totalRho}) for the single quench case].
Explicitly written out, these are given by
\begin{align}
\tilde{\rho}&^{++}_{rs}(m,\tilde{\tau}_p)\nonumber\\
=&\sum_{m_1=m+1}^N\sum_{l_1e_1e}{_{Q_{p+1}}}\langle rem|e^{-iH^{Q_p}{\tau}_p}...e^{-iH^{Q_1}{\tau}_1}|l_1e_1m_1\rangle{_i}\nonumber\\
&\times w_{m_1} \frac{e^{-\beta E_{l_1}^{m_1}}}{\tilde{Z}_{m_1}}{_i}\langle l_1e_1m_1|e^{iH^{Q_1}{\tau}_1}...e^{iH^{Q_p}{\tau}_p}| sem\rangle_{Q_{p+1}}\nonumber\\
=&\sum_{m_1=m+1}^N\sum_{l_1e_1e}{_{Q_{p+1}}}\langle rem|e^{-iH^{Q_p}{\tau}_p}...e^{-iH^{Q_1}{\tau}_1}(1^{i+}_m+1^{i-}_m)|l_1e_1m_1\rangle{_i}\nonumber\\
&\times w_{m_1} \frac{e^{-\beta E_{l_1}^{m_1}}}{\tilde{Z}_{m_1}}{_i}\langle l_1e_1m_1|(1^{i+}_m+1^{i-}_m)e^{iH^{Q_1}{\tau}_1}...e^{iH^{Q_p}{\tau}_p}| sem\rangle_{Q_{p+1}}\nonumber\\
=&\sum_{kk'ee'e''}{_{Q_{p+1}}}\langle rem|e^{-iH^{Q_p}{\tau}_p}...e^{-iH^{Q_1}{\tau}_1}|ke'm\rangle{_i}\nonumber\\
&\times \sum_{m_1=m+1}^N\sum_{l_1e_1}{_i}\langle ke'm|l_1e_1m_1\rangle{_i}w_{m_1} \frac{e^{-\beta E_{l_1}^{m_1}}}{\tilde{Z}_{m_1}}{_i}\langle l_1e_1m_1|k'e''m\rangle{_i}\nonumber\\
&\times{_i}\langle k'e''m|e^{iH^{Q_1}{\tau}_1}...e^{iH^{Q_p}{\tau}_p}| sem\rangle_{Q_{p+1}},\label{eq:rho++generic}\end{align}
\begin{align}
\rho^{0}_{rs}(m,\tilde{\tau}_p)=&\sum_{lee_1}{_{Q_{p+1}}}\langle rem|e^{-iH^{Q_p}{\tau}_p}...e^{-iH^{Q_1}{\tau}_1}|le_1m\rangle{_i}w_{m} \frac{e^{-\beta E_{l}^{m}}}{\tilde{Z}_{m}}\nonumber\\
&\times {_i}\langle le_1m|e^{iH^{Q_1}{\tau}_1}...e^{iH^{Q_p}{\tau}_p}| sem\rangle_{Q_{p+1}},\label{eq:rho0generic}
\end{align}
and,
\begin{align}
\rho&^{--}_{rs}(m,\tilde{\tau}_p)\nonumber\\
=&\sum_{m_1=m_0}^{m-1}\sum_{l_1e_1e}{_{Q_{p+1}}}\langle rem|e^{-iH^{Q_p}{\tau}_p}...e^{-iH^{Q_1}{\tau}_1}|l_1e_1m_1\rangle{_i}\nonumber\\&\times w_{m_1} \frac{e^{-\beta E_{l_1}^{m_1}}}{\tilde{Z}_{m_1}}{_i}\langle l_1e_1m_1|e^{iH^{Q_1}{\tau}_1}...e^{iH^{Q_p}{\tau}_p}| sem\rangle_{Q_{p+1}}\nonumber\\
=&\sum_{m_1=m_0}^{m-1}\sum_{l_1e_1e}{_{Q_{p+1}}}\langle rem|(1^{Q_{p+1}+}_{m_1}+1^{Q_{p+1}-}_{m_1})e^{-iH^{Q_p}{\tau}_p}...e^{-iH^{Q_1}{\tau}_1}|l_1e_1m_1\rangle{_i}\nonumber\\&\times w_{m_1} \frac{e^{-\beta E_{l_1}^{m_1}}}{\tilde{Z}_{m_1}}{_i}\langle l_1e_1m_1|e^{iH^{Q_1}{\tau}_1}...e^{iH^{Q_p}{\tau}_p}(1^{Q_{p+1}+}_{m_1}+1^{Q_{p+1}-}_{m_1})| sem\rangle_{Q_{p+1}}\nonumber\\
=&\sum_{kk'ee'e''}\sum_{m_1=m_0}^{m-1}\sum_{l_1e_1}{_{Q_{p+1}}}\langle rem|ke'm_1\rangle{_{Q_{p+1}}}\nonumber\\
&\times {_{Q_{p+1}}}\langle ke'm_1|e^{-iH^{Q_p}{\tau}_p}...e^{-iH^{Q_1}{\tau}_1}|l_1e_1m_1\rangle{_i}w_{m_1} \frac{e^{-\beta E_{l_1}^{m_1}}}{\tilde{Z}_{m_1}}\nonumber\\
&\times{_i}\langle l_1e_1m_1|e^{iH^{Q_1}{\tau}_1}...e^{iH^{Q_p}{\tau}_p}|k'e''m_1\rangle{_{Q_{p+1}}}{_{Q_{p+1}}}\langle k'e''m_1| sem\rangle_{Q_{p+1}}\label{eq:rho--generic},
\end{align}
in which we used $1^{i-}_m|l_1e_1m_1\rangle{_i}=0$ for $m_1>m$, $1^{Q_{p+1}-}_{m_1}| sem\rangle_{Q_{p+1}}=0$ for $m_1<m$, 
and $1^+_m=\sum_{ke}|kem\rangle \langle kem|$. Noting that the matrix element, ${_{Q_{p+1}}}\langle rem|e^{-iH^{Q_p}{\tau}_p}...e^{-iH^{Q_1}{\tau}_1}|se'm\rangle{_i}$, appears in all three terms above, we denote this matrix element by
\begin{align}
\mathcal{S}^{m}_{r_{Q_{p+1}}s_i}(ee',-\tilde{\tau}_p)={_{Q_{p+1}}}\langle rem|e^{-iH^{Q_p}{\tau}_p}...e^{-iH^{Q_1}{\tau}_1}|se'm\rangle{_i}\label{overlapTime},
\end{align}
and call it the generalized overlap matrix element.
\footnote{
The factor $e^{-iH^{Q_p}{\tau}_p}\dots e^{-iH^{Q_1}{\tau}_1}$ is the time evolution operator
at time $\tilde{\tau}_{p}=\sum_{i=1}^{p}\tau_i$ following $p$ intermediate quantum quenches  described by $H^{Q_1},\dots,H^{Q_{p}}$. 
Hence, $\protect{{_{Q_{p+1}}}\langle rem|e^{-iH^{Q_p}{\tau}_p}\dots e^{-iH^{Q_1}{\tau}_1}|se'm\rangle{_i}}$
is the matrix element of this operator between the initial states of $H^{i}$ and the states of the quench 
Hamiltonian $H^{Q_{p+1}}$.
}
We can prove, see Appendix~\ref{sec:tau-overlap-matrices}, that these generalized overlap matrix elements are diagonal in the environment variables, 
\begin{align}
\mathcal{S}^m_{r_{Q_{p+1}}s_i}(ee',-\tilde{\tau}_p)&=\mathcal{S}^m_{r_{Q_{p+1}}s_i}(-\tilde{\tau}_p)\delta_{ee'}\label{eq:totalStau},
\end{align}
with $\mathcal{S}^m_{r_{Q_{p+1}}s_i}(-\tilde{\tau}_p)$ being decomposed into three terms
\begin{align}
&\mathcal{S}^m_{r_{Q_{p+1}}s_i}(-\tilde{\tau}_p)=\mathcal{S}^{m++}_{r_{Q_{p+1}}s_i}(-\tilde{\tau}_p)+\mathcal{S}^{m0}_{r_{Q_{p+1}}s_i}(-\tilde{\tau}_p)+\mathcal{S}^{m--}_{r_{Q_{p+1}}s_i}(-\tilde{\tau}_p)
\end{align}
These terms are determined by the following equations,
\begin{align}
&\mathcal{S}^{m++}_{r_{Q_{p+1}}s_i}(-\tilde{\tau}_p)=\sum_{k}S^m_{r_{Q_{p+1}}k_{Q_p}}e^{-iE^m_k{\tau}_p}\mathcal{S}^m_{k_{Q_p}s_i}(-\tilde{\tau}_{p-1})\label{eq:totalStau++}\\
&\mathcal{S}^{m0}_{r_{Q_{p+1}}s_i}(-\tilde{\tau}_p)=\sum_{l}S^m_{r_{Q_{p+1}}l_{Q_p}}e^{-iE^m_l{\tau}_p}\mathcal{S}^m_{l_{Q_p}s_i}(-\tilde{\tau}_{p-1})\label{eq:totalStau0}
\\
&\mathcal{S}^{m--}_{r_{Q_{p+1}}s_i}(-\tilde{\tau}_p)
=\sum_{\alpha_m}\sum_{kk'}A^{\alpha_m\dagger}_{rk}\Big[\mathcal{S}^{(m-1)0}_{k_{Q_{p+1}}k'_i}(-\tilde{\tau}_{p})+\mathcal{S}^{(m-1)--}_{k_{Q_{p+1}}k'_i}(-\tilde{\tau}_{p})\Big]A^{\alpha_m}_{k's}\label{eq:totalStau--}\nonumber\\
&\hspace{6em}\text{with\quad } \mathcal{S}^{m_0--}_{r_{Q_{p+1}}s_i}(-\tilde{\tau}_p)=0.\end{align}
In the above, and in subsequent equations, indices such as $k_{Q_{p+1}}k_{i}'$ appearing in full within summands, are abbreviated
to $kk'$ in summation subscripts (as in $\sum_{kk'}$ above).
The recursion relations given by Eqs.~(\ref{eq:totalStau})-(\ref{eq:totalStau--}) allow the generalized overlap matrix elements in Eq.~(\ref{eq:totalStau}) to be calculated 
at an arbitrary time step, $\tilde{\tau}_p>\tilde{\tau}_1$. While $\mathcal{S}^{m++}$ and $\mathcal{S}^{m0}$ depend on the generalized overlap matrix element at one earlier time step, $\mathcal{S}^{m--}$ depends on the $0$ and $--$ components of these matrix elements at the same time step, but at one earlier shell. 
For a detailed proof of the above equations as well as the closed form of the generalized overlap matrix elements at $\tilde{\tau}_1$ we refer the reader to Appendix~\ref{sec:tau-overlap-matrices}.

Returning to the terms in the projected density matrix in Eqs.~(\ref{eq:rho++generic})-(\ref{eq:rho--generic}), we substitute the generalized overlap matrix element, which is diagonal in the environment variable, into each term as follows
\begin{align}
\tilde{\rho}^{++}_{rs}(m,\tilde{\tau}_p)=&\sum_{kk'}\mathcal{S}^m_{r_{Q_{p+1}}k_i}(-\tilde{\tau}_p)R^m_{\rm red}(k,k')\mathcal{S}^m_{k'_is_{Q_{p+1}}}(\tilde{\tau}_p)\label{eq:rho++multiple}\\
\rho^{0}_{rs}(m,\tilde{\tau}_p)=&\sum_{l}\mathcal{S}^m_{r_{Q_{p+1}}l_i}(-\tilde{\tau}_p)w_{m} \frac{e^{-\beta E_{l}^{m}}}{{Z}_{m}}\mathcal{S}^m_{l_is_{Q_{p+1}}}(\tilde{\tau}_p)\label{eq:rho0multiple}\\
\rho^{--}_{rs}(m,\tilde{\tau}_p)=&\sum_{kk'e}\sum_{m_1=m_0}^{m-1}\sum_{l_1e_1}{_{Q_{p+1}}}\langle rem|ke_1m_1\rangle{_{Q_{p+1}}}\mathcal{S}^{m_1}_{k_{Q_{p+1}}l_{1i}}(-\tilde{\tau}_p)\nonumber\\
&\hspace{2em}\times w_{m_1} \frac{e^{-\beta E_{l_1}^{m_1}}}{\tilde{Z}_{m_1}}\mathcal{S}^{m_1}_{l_{1i}k'_{Q_{p+1}}}(\tilde{\tau}_p){_{Q_{p+1}}}\langle k'e_1m_1| sem\rangle_{Q_{p+1}}\nonumber\\
=&\sum_{kk'e}\sum_{m_1=m_0}^{m-1}\sum_{e_1}{_{Q_{p+1}}}\langle rem|ke_1m_1\rangle{_{Q_{p+1}}}\frac{1}{d^{N-m_1}}\rho^0_{kk'}(m_1,\tilde{\tau}_p)\nonumber\\
&\hspace{8em}\times{_{Q_{p+1}}}\langle k'e_1m_1| sem\rangle_{Q_{p+1}}\nonumber\\
=&\frac{1}{d}\sum_{kk'\alpha_{m}}A^{\alpha_{m}\dagger}_{rk}\bigg\{\rho^0_{kk'}(m-1,\tilde{\tau}_p)+\rho^{--}_{kk'}(m-1,\tilde{\tau}_p)\bigg\}A^{\alpha_{m}}_{k's}\nonumber\\
& \text{with\quad } \rho^{--}_{rs}(m_0,\tilde{\tau}_p)=0\label{eq:rho--multiple},
\end{align}
in which $\mathcal{S}^m_{r_is_{Q_{p+1}}}(\tilde{\tau}_p)=[\mathcal{S}^m_{s_{Q_{p+1}}r_i}(-\tilde{\tau}_p)]^{\dagger}$, $R^m_{\rm red}(k,k')$ is defined as in Eq.~(\ref{eq:RFDM}), 
and the recursion relation for $\rho^{--}_{rs}(m,\tilde{\tau}_p)$ in Eq.~(\ref{eq:rho--multiple}) is derived as for $\rho^{--}_{rs}(m)$ in section~\ref{sec:section-3}. 
The above equations determine the projected density matrix of Eq.~(\ref{eq:Totalrhomultiquenches}) at an arbitrary time step $\tilde{\tau}_i\geq\tilde{\tau}_1$. 

One can use the scheme in previous works\cite{Anders2005,Anders2006} to calculate the projected density matrix for multiple quenches, which yield a simplified $\rho^{i\to Q_{p+1}}_{rs}(m,\tilde{\tau}_p)=\tilde{\rho}^{++}_{rs}(m,\tilde{\tau}_p)+\rho^{0}_{rs}(m,\tilde{\tau}_p)$ since $\rho^{--}_{rs}(m,\tilde{\tau}_p)=0$ can be neglected at any time step. 
If one also neglects the term $\mathcal{S}^{m--}_{r_is_{Q_{p+1}}}(\tilde{\tau}_p)$ of the generalized overlap matrix element, presented above, then $\mathcal{S}^{m}_{r_is_{Q_{p+1}}}(\tilde{\tau}_p)\approx\mathcal{S}^{m++}_{r_is_{Q_{p+1}}}(\tilde{\tau}_p)+\mathcal{S}^{m0}_{r_is_{Q_{p+1}}}(\tilde{\tau}_p)$. Using the definitions in Eqs.~(\ref{eq:totalStau++})-(\ref{eq:totalStau0}), we have
\begin{align}
\rho^{i\to Q_{p+1}}_{rs}(m,\tilde{\tau}_p)\approx\sum_{qq'}S^m_{r_{Q_{p+1}}q_{Q_p}}e^{-i(E^m_q-E^m_{q'})\tau_p}\rho^{i\to Q_{p}}_{qq'}(m,\tilde{\tau}_{p-1})S^m_{q'_{Q_p}s_{Q_{p+1}}}.\nonumber
\end{align}
This equation is similar to Eqs.~(56)-(57) in the recent work of Ref.~\onlinecite{Eidelstein2012}, which investigates time evolution due to periodic switching. However, $\mathcal{S}^{m--}_{r_is_{Q_{p+1}}}(\tilde{\tau}_p)$ equals zero only at $m=m_0$. Omitting this term leads to an accumulated error after each time step due to the causality 
in Eqs.~(\ref{eq:totalStau})-(\ref{eq:totalStau--}), where generalized overlap matrix elements at a given time step are shown to depend on those at one earlier time step. This omission may be equivalent to the approximation  in the mentioned work, where an accumulated error is observed in the time evolution.

In general, by following the procedure for calculating the projected density matrix for the single quench case in Sec.~\ref{sec:section-2}, and especially the term $\rho^{--}_{rs}(m)$ in Sec.~\ref{sec:section-3}, we explicitly formulated all the terms of the projected density matrix in Eqs.~(\ref{eq:rho++multiple})-(\ref{eq:rho--multiple}) for the
multiple quench case. Substituting this projected density matrix into Eq.~(\ref{eq:Otime}), one can estimate the time evolution of a local observable at
an arbitrary time, $t\geq\tilde{\tau}_1$. For $0=\tilde{\tau}_0\leq t<\tilde{\tau}_1$, one recovers the TDNRG with a single quench as in Secs~\ref{sec:section-1}-\ref{sec:section-3}. The entire procedure of generalizing the TDNRG to an arbitrary number of quenches, presented here, adopts no further approximation than does the TDNRG with a single quench. Therefore, by ensuring sufficiently small quantum quenches, one can expect, by the arguments presented
in the introduction to this section, to obtain high accuracy for the time evolution even in the limit of long times. 
The numerical implementation of this lies outside the scope of the present paper and will be reported elsewhere, however, a few remarks about the
computational cost of such an approach are in order. In comparison to the single quench case, the main increase in computational cost comes from, (a),
the requirement to calculate the projected density matrix $\rho^{i\to Q_{p+1}}_{sr}(m,\tilde{\tau}_p)$ for $p=1,\dots,n$, i.e. for $n$ quenches  instead of a single quench,
\footnote{For times $t>\tilde{\tau}_n$. At shorter times,  $\tilde{\tau}_{p+1}>t\geq\tilde{\tau}_p$, with $p<n$, one needs to 
calculate only the first $p$ projected density matrices, so the computational cost becomes less for shorter times.} 
and, (b),  the requirement to calculate the generalized overlap matrix elements, $\mathcal{S}^m_{r_{Q_{p+1}}s_i}(-\tilde{\tau}_p)$, 
also for $n$ quenches, using an algorithm which resembles that used for the calculation of reduced density matrices. 
Since the latter are implemented highly efficiently,  we estimate the computational cost to be approximately linear in the number of quenches and therefore feasible.
\section{Conclusions and Outlook}
\label{sec:conclusions}
In this paper, we generalized the TDNRG method for describing the time evolution of a local observable 
following a single quantum quench to arbitrary finite temperature, starting from the full density matrix 
of the initial state. 
In this generalization, we made no further approximation than the original NRG approximation [Eq.~(\ref{eq:nrg-approx})]. 
The generalization relies on determining the projected density matrix, which is made up of three parts $\tilde{\rho}^{++}, \rho^{--}$ 
and $\rho^{0}$, for which explicit expressions, and recursion relations for their evaluation, were obtained. 
The trace conserving property of the projected density matrix was proven and was used to check the numerical 
precision of the calculations.  We find that the contribution from $\rho^{--}$, absent in previous approaches,\cite{Anders2005,Anders2008} 
is significant at finite temperature and cannot be neglected.
The short-time limit of local observables was shown to be exactly equal to the thermodynamic value in the initial state, 
both analytically and numerically (Sec.~\ref{subsec:short-time-limit}). 
Furthermore, we clarified that the NRG approximation is the main source of noise in local observables 
at long times (Sec.~\ref{subsec:time-dependence}), whereas it appears not responsible for the error in the long-time limit.
The latter is found to be reduced by either reducing $\Lambda$ (i.e. approaching the continuum limit $\Lambda\to 1^{+}$), 
as noted previously\cite{Anders2006}, or by reducing the size of the quantum quench (Sec.~\ref{subsec:long-time-limit}). 
Also, the short-time behavior is improved with decreasing quench size, see Fig.~\ref{fig:tdNRGvsAnalytic}.

The above results suggested, since the limit $\Lambda\to 1^+$ is impractical in NRG calculations, a formulation in which a large quench is
replaced by multiple quenches over a finite time interval as a means to improve the TDNRG at all time scales, and particularly
for longer times. We provided this multiple quench formulation (Sec.~\ref{sec:generic}) and showed that, as for the
single quench case, it rests solely on a single approximation (the NRG approximation). The structure of this theory, while formally 
similar to the single quench case, brings out more clearly the importance of including the ``$--$'' terms, i.e., $\rho^{--}$ and $\mathcal{S}^{m--}$. Namely, neglecting $\mathcal{S}^{m--}$
in the multiple quench case leads to an accumulated error after each time step $\tilde{\tau}_p, p=1,\dots,n$ due to the causality
of Eqs.~(\ref{eq:totalStau})-(\ref{eq:totalStau--}), hence, it will be particularly important not to approximate this in a future 
numerical implementation of the multiple-quench TDNRG. The latter, as discussed at the end of the previous section is feasible, 
scaling at most linearly with the number of quenches, but lies beyond the scope of the present paper and will be presented elsewhere. 
Our formalism for the multiple quench case  also generalizes the TDNRG to general pulses and periodic switching/driving. The latter 
has been considered also in Ref.~\onlinecite{Eidelstein2012}, within a hybrid TDNRG DMRG approach, although several approximations 
in addition to the NRG approximation were made (see the discussion in Sec.~\ref{sec:generic}).

We also compared our single-quench TDNRG approach to the previous scheme,\cite{Anders2005,Anders2006} finding that the present 
one gives results with less noise at long times, with the results being closer to the expected long-time limit, 
particularly at finite temperatures. This is due to the contribution of the full Wilson chain in our TDNRG using the
full density matrix of the initial state, which at finite temperature always results in a finite contribution from $\rho^{--}$, 
absent in the scheme of Refs.~\onlinecite{Anders2005,Anders2006} which uses a truncated Wilson chain for finite temperatures.

In future, it would be of interest, especially in the light of time-resolved experiments, e.g., time-resolved photoemission and scanning tunneling microscope spectroscopies,
\cite{Perfetti2006,Loth2010} to extend the formalism in this paper to the time-evolution of 
dynamical quantities following a quench/pulse, e.g., to single-particle spectral functions $A(\omega,t)$,\cite{Iyoda2013,Freericks2009}
%
%
dynamical spin susceptibilities,\cite{Medvedyeva2013} or optical conductivities. \cite{Eckstein2008}
This would allow basic questions about the time-dependence of the Kondo effect in quantum dots to be addressed, e.g. the time-evolution of the Kondo resonance upon instantaneously 
switching from the mixed-valence to the Kondo regime,\cite{Nordlander1999} or, the transient response of the current following a bias voltage pulse of finite duration.\cite{Plihal2000} 
Given the common matrix product state structure of the NRG and DMRG eigenstates,\cite{Verstraete2008} some of the formalism developed in this paper for general multiple quenches might find application also within the time-dependent DMRG. Finally, improved TDNRG approaches would be of interest in the development of methods for studying steady-state non-equilibrium transport through correlated systems, such as quantum dots.\cite{Anders2008} 
\begin{acknowledgments}
We thank D.~P. DiVincenzo,  S. Andergassen, V. Meden, A. Weichselbaum and F.~B. Anders for useful discussions and encouraging remarks, 
and acknowledge supercomputer support by the John von Neumann institute for Computing (J\"ulich).
\end{acknowledgments}
\appendix
\section{Trace of the Projected Density Matrix}
\label{sec:trace-formula}
Starting from the exact Eq.~(\ref{eq:Ot}) and setting $\hat{O}=1$ results in $1={\rm Tr}[\rho]$, a tautology, since $\rho$ is normalized.
However, the result we are interested in starts  from Eq.~(\ref{eq:localOt}), which in general will be approximate due to the use of the
NRG approximation. Inserting $\hat{O}=1$ in the right-hand side (RHS) of this expression and using $O^m_{rs}=\delta_{rs}$ results in Eq.~(\ref{eq:identity})
which states that the trace of the projected density matrix is preserved. The result is exact, for the same reason that the short-time limit is 
exact (as discussed in Sec.~\ref{subsec:exact results}), namely any error that would arise from approximate NRG eigenvalues in 
Eq.~(\ref{eq:localOt}), does not arise for $\hat{O}=1$ because these eigenvalues drop out  from the expression (\ref{eq:localOt}) for this choice
of operator. More explicitly, using the definition of the FDM, $\rho=\sum_{m=m_{0}}^Nw_m\tilde{\rho}_m$, 
from Sec.~\ref{subsec:nrg+cbs}  and the completeness relation $\sum_{m=m_0}^N \sum_{le} | lem\rangle\langle lem|=1$, we have for the RHS of Eq.~(\ref{eq:identity}),
\begin{align}
&\sum_{m=m_0}^{N}\sum_{l}\rho_{ll}^{i\to f}(m)\equiv \sum_{m=m_0}^N \sum_{l,e}{_f}\langle lem|\rho | lem\rangle_f\nonumber\\
=&\sum_{m=m_0}^N \sum_{m'=m_0}^N w_{m'}\sum_{l,e}\sum_{l',e'}{_f}\langle lem|l'e'm'\rangle{_i} \frac{e^{-\beta E_{l'}^{m'}}}{\tilde{Z}_{m'}}{_i}\langle l'e'm' | lem\rangle_f\nonumber\\
=&\sum_{m=m_0}^N \sum_{m'=m_0}^N w_{m'}\sum_{l,e}\sum_{l',e'}{_i}\langle l'e'm' | lem\rangle_f{_f}\langle lem|l'e'm'\rangle{_i} \frac{e^{-\beta E_{l'}^{m'}}}{\tilde{Z}_{m'}}\nonumber\\
=&\sum_{m'=m_0}^N w_{m'}\sum_{l',e'}{_i}\langle l'e'm' |l'e'm'\rangle{_i} \frac{e^{-\beta E_{l'}^{m'}}}{\tilde{Z}_{m'}}\nonumber\\
=&\sum_{m'=m_0}^N w_{m'}\sum_{l'}\frac{e^{-\beta E_{l'}^{m'}}}{{Z}_{m'}}=1\label{eq:identity2},
\end{align}
where, in the last line, $\sum_{m'=m_0}^{N}w_{m'}\equiv {\rm Tr}[\rho]=1$ was used. This verifies Eq.~(\ref{eq:identity}), which serves as a useful identity to check that, (a), the  three terms contributing to $\rho^{i\to f}(m)$
in Eq.~(\ref{eq:totalRho}) have been evaluated correctly, and, (b), the numerical calculation of $\rho^{i\to f}(m)$ is sufficient. This was the case, since we found an error of less than $10^{-10}$ for all 
$\Lambda$ and all system parameters used in the Anderson model  ($\varepsilon_{d}$, $U$); see Fig.~\ref{fig:traceRho}.
\section{Short-Time Limit}
\label{sec:appendix-short-time-limit}
In this appendix we show explicitly that the TDNRG expression for the expectation value of a local observable 
in the short-time limit ($t\to 0^{+}$) recovers the correct thermodynamic value in the initial state, i.e. that
 \begin{align}
 O(t\to 0^{+})\equiv\sum_{m=m_0}^N \sum_{rs\notin KK'}\rho^{i\to f}_{sr}(m) O^m_{rs}= O_i \label{eq:Ot0appendix}.
 \end{align}
 where $O_{i}={\rm Tr}\left[\rho \hat{O}\right]$.

Similar to the calculation of the trace of the projected density matrix in Eq.~(\ref{eq:identity2}), we substitute $\rho=\sum_{m=m_{0}}^Nw_m\tilde{\rho}_m$  
into Eq. (\ref{eq:Ot0appendix}), obtaining
\begin{align}
O(t\to 0^{+})=&\sum_{m=m_0}^N \sum_{rse}^{\ne KK'}{_f}\langle rem|\rho | sem\rangle_f  {_f}\langle sem| \hat{O}| rem\rangle_f\nonumber\\
=&\sum_{m=m_0}^N \sum_{rse}^{\ne KK'}\sum_{m'=m_0}^N  \sum_{l'e'}{_f}\langle rem|l'e'm'\rangle{_i}w_{m'} \frac{e^{-\beta E_{l'}^{m'}}}{\tilde{Z}_{m'}}\nonumber\\
&\times{_i}\langle l'e'm' | sem\rangle_f  {_f}\langle sem| \hat{O}| rem\rangle_f\nonumber\\
=&\sum_{m'=m_0}^N\sum_{l'e'} w_{m'} \frac{e^{-\beta E_{l'}^{m'}}}{\tilde{Z}_{m'}}\Big\{\sum_{m=m_0}^N \sum_{rse}^{\ne KK'}{_i}\langle l'e'm' | sem\rangle_f \nonumber\\
&\times {_f}\langle sem| \hat{O}| rem\rangle_f{_f}\langle rem|l'e'm'\rangle{_i}\Big\} \label{eq:Otfull2}.\end{align}
Decomposing $| rem\rangle_f$ into kept, $| kem\rangle_f$, and discarded, $| lem\rangle_f$, states, we obtain for the part inside the curly brackets
\begin{align}
&\sum_{m=m_0}^N \sum_{lse}{_i}\langle l'e'm' | sem\rangle_f  {_f}\langle sem| \hat{O}| lem\rangle_f{_f}\langle lem|l'e'm'\rangle{_i}\nonumber\\
+&\sum_{m=m_0}^N\sum_{kle}{_i}\langle l'e'm' | lem\rangle_f  {_f}\langle lem| \hat{O}| kem\rangle_f{_f}\langle kem|l'e'm'\rangle{_i}\nonumber\\
=&\sum_{m=m_0}^N \sum_{m_1=m}^N\sum_{le}\sum_{l_1e_1}{_i}\langle l'e'm' | l_1e_1m_1\rangle_f  {_f}\langle l_1e_1m_1| \hat{O}| lem\rangle_f\nonumber\\
&\hspace{14em}\times{_f}\langle lem|l'e'm'\rangle{_i}\nonumber\\
+&\sum_{m=m_0}^N\sum_{m_1=m+1}^N\sum_{l_1e_1}\sum_{le}{_i}\langle l'e'm' | lem\rangle_f  {_f}\langle lem| \hat{O}| l_1e_1m_1\rangle_f\nonumber\\
&\hspace{12em}\times{_f}\langle l_1e_1m_1|l'e'm'\rangle{_i}\label{eq:curlypart}.
\end{align}
Replacing $\sum_{m=m_0}^N \sum_{m_1=m}^N\to\sum_{m_1=m_0}^N \sum_{m=m_0}^{m_1}$ in the first term, and interchanging $| lem\rangle\leftrightarrow| l_1e_1m_1\rangle$ in the second one, we have Eq.~(\ref{eq:curlypart}) equal to
\begin{align}
&\sum_{m_1=m_0}^N \sum_{m=m_0}^{m_1}\sum_{le}\sum_{l_1e_1}{_i}\langle l'e'm' | l_1e_1m_1\rangle_f  {_f}\langle l_1e_1m_1| \hat{O}| lem\rangle_f\nonumber\\
&\hspace{14em}\times{_f}\langle lem|l'e'm'\rangle{_i}\nonumber\\
+&\sum_{m_1=m_0}^N\sum_{m=m+1}^N\sum_{le}\sum_{l_1e_1}{_i}\langle l'e'm' | l_1e_1m_1\rangle_f  {_f}\langle l_1e_1m_1| \hat{O}| lem\rangle_f\nonumber\\
&\hspace{14em}\times{_f}\langle lem|l'e'm'\rangle{_i}\nonumber\\
=&\sum_{m_1=m_0}^N \sum_{m=m_0}^{N}\sum_{le}\sum_{l_1e_1}{_i}\langle l'e'm' | l_1e_1m_1\rangle_f  {_f}\langle l_1e_1m_1| \hat{O}| lem\rangle_f\nonumber\\
&\hspace{14em}\times{_f}\langle lem|l'e'm'\rangle{_i}\nonumber\\
=&{_i}\langle l'e'm' |\hat{O}|l'e'm'\rangle{_i}
\label{eq:Of2i}.
\end{align}
Substituting Eq.~(\ref{eq:Of2i}) into Eq.~(\ref{eq:Otfull2}), we have 
\begin{align}
O(t\to 0^{+})=&\sum_{m'=m_0}^N\sum_{l'e'} w_{m'} \frac{e^{-\beta E_{l'}^{m'}}}{\tilde{Z}_{m'}}{_i}\langle l'e'm' |\hat{O}|l'e'm'\rangle{_i}\nonumber\\
=&\sum_{m'=m_0}^N\sum_{l'} w_{m'} \frac{e^{-\beta E_{l'}^{m'}}}{{Z}_{m'}}O_{l'l'} \label{eq:Ot0full},\end{align}
which is the thermodynamic average of the observable in the initial state, as calculated within the FDM approach. This proves that $O(t\to 0^{+})=O_i$ is always true. 
We also verified the latter numerically, for the Anderson impurity model, finding that the calculated value of $O(t\to 0^{+})$ lies within $10^{-10}$ of $O_i$ regardless of 
the parameter sets used and regardless of the discretization parameter $\Lambda$.
\section{Recursion Relations for SU(2) Symmetry}
\label{sec:su2}
In the actual calculations for the Anderson impurity model, presented in this paper, SU(2) spin symmetry was used in the numerical diagonalizations. Accordingly, the states were classified according to their values of total charge $Q$, total spin $S$, and $z$ component of total spin $S_z$\cite{Wilson1975}. 
Moreover, as the eigenvalues are independent of $S_z$, we need only write the recursion relations in the $(Q,S)$ subspace. The transformation matrix $A_{kr}^{\alpha_{m}}$ in Eq.~(\ref{eq:transformation-matrix}) is now denoted by $U^{m}(ki,r)$  as in the original work in Ref.~\onlinecite{KWW1980a}, 
with the correspondence $i\leftrightarrow \alpha_{m}$. In detail, $\rho^{--}_{sr}(m)$ is expressed in terms of the following four terms, corresponding to $i$ taking one of four values ($d=4$).
\begin{align}
&\rho_{sr}^{--}(Q,S,m)\nonumber\\
=&\frac{1}{4}\sum_{k,k'}U^{m\dagger}_{Q,S}(k1,s)\tilde{\rho}_{k,k'}(Q+1,S,m-1)U^{m}_{Q,S}(k'1,r)\nonumber\\
+&\frac{1}{4}\sum_{k,k'}U^{m\dagger}_{Q,S}(k2,s)\tilde{\rho}_{k,k'}(Q,S-\frac{1}{2},m-1)U^{m}_{Q,S}(k'2,r)\nonumber\\
+&\frac{1}{4}\sum_{k,k'}U^{m\dagger}_{Q,S}(k3,s)\tilde{\rho}_{k,k'}(Q,S+\frac{1}{2},m-1)U^{m}_{Q,S}(k'3,r)\nonumber\\
+&\frac{1}{4}\sum_{k,k'}U^{m\dagger}_{Q,S}(k4,s)\tilde{\rho}_{k,k'}(Q-1,S,m-1)U^{m}_{Q,S}(k'4,r)
\label{eq:rho--inQS},
\end{align}
with $\tilde{\rho}_{k,k'}(Q,S,m)=\rho_{k,k'}^{0}(Q,S,m)+\rho_{k,k'}^{--}(Q,S,m)$. 

Similarly, the recursion relation for the reduced full density matrices takes the form
\begin{align}
&[R_{\rm red}^{m}(Q,S)]_{k,k'}\nonumber\\
=&\sum_{q,q'}U^{m+1}_{Q-1,S}(k1,q)[\tilde{R}_{\rm red}^{m+1}(Q-1,S)]_{q,q'}U^{m+1\dagger}_{Q-1,S}(k'1,q')\nonumber\\
+&\frac{2S+2}{2S+1}\sum_{q,q'}U^{m+1}_{Q,S+\frac{1}{2}}(k2,q)[\tilde{R}_{\rm red}^{m+1}(Q,S+\frac{1}{2})]_{q,q'}U^{m+1\dagger}_{Q,S+\frac{1}{2}}(k'2,q')\nonumber\\
+&\frac{2S}{2S+1}\sum_{q,q'}U^{m+1}_{Q,S-\frac{1}{2}}(k3,q)[\tilde{R}_{\rm red}^{m+1}(Q,S-\frac{1}{2})]_{q,q'}U^{m+1\dagger}_{Q,S-\frac{1}{2}}(k'3,q')\nonumber\\
+&\sum_{q,q'}U^{m+1}_{Q+1,S}(k4,q)[\tilde{R}_{\rm red}^{m+1}(Q+1,S)]_{q,q'}U^{m+1\dagger}_{Q+1,S}(k'4,q')\label{eq:RinQS}.
\end{align}
with $\begin{bmatrix} \tilde{R}_{\rm red}^{m}(Q,S) \end{bmatrix} =\begin{bmatrix} \Big[{R}_{\rm red}^{m}(Q,S)\Big]_{k,k'}& 0 & \cdots &0 \\ 0 &w_{{m}}\frac{e^{-\beta E_{Q,S,l}^{m}}}{{Z}_{m}}&\cdots & 0 \\ \vdots&\vdots &\ddots &\vdots\\0 & 0 &\cdots & w_{{m}}\frac{e^{-\beta E_{Q,S,l}^{m}}}{{Z}_{m}}\end{bmatrix}$, and noting  that $\sum_q=\sum_k+\sum_l$ and $|qm\rangle=|km\rangle+|lm\rangle$.
\section{Long-Time Limit: Additional Switching Protocols}
\label{sec:supporting}
In this appendix, we present additional results for the long time limit of observables with different switching protocols to support the conclusions in Sec.~\ref{subsec:long-time-limit}. 
In general, the results show us the trend of quench size dependence as observed in Fig.~\ref{fig:Otinf}. Depending on the local observable, i.e., local level occupation number or double occupancy, the results depend on different parameters of the final state.

\subsection{Double occupancy: switching  $\varepsilon_d$ with $U$ constant}
\begin{figure}[h]
\centering
\includegraphics[width=0.51\textwidth]{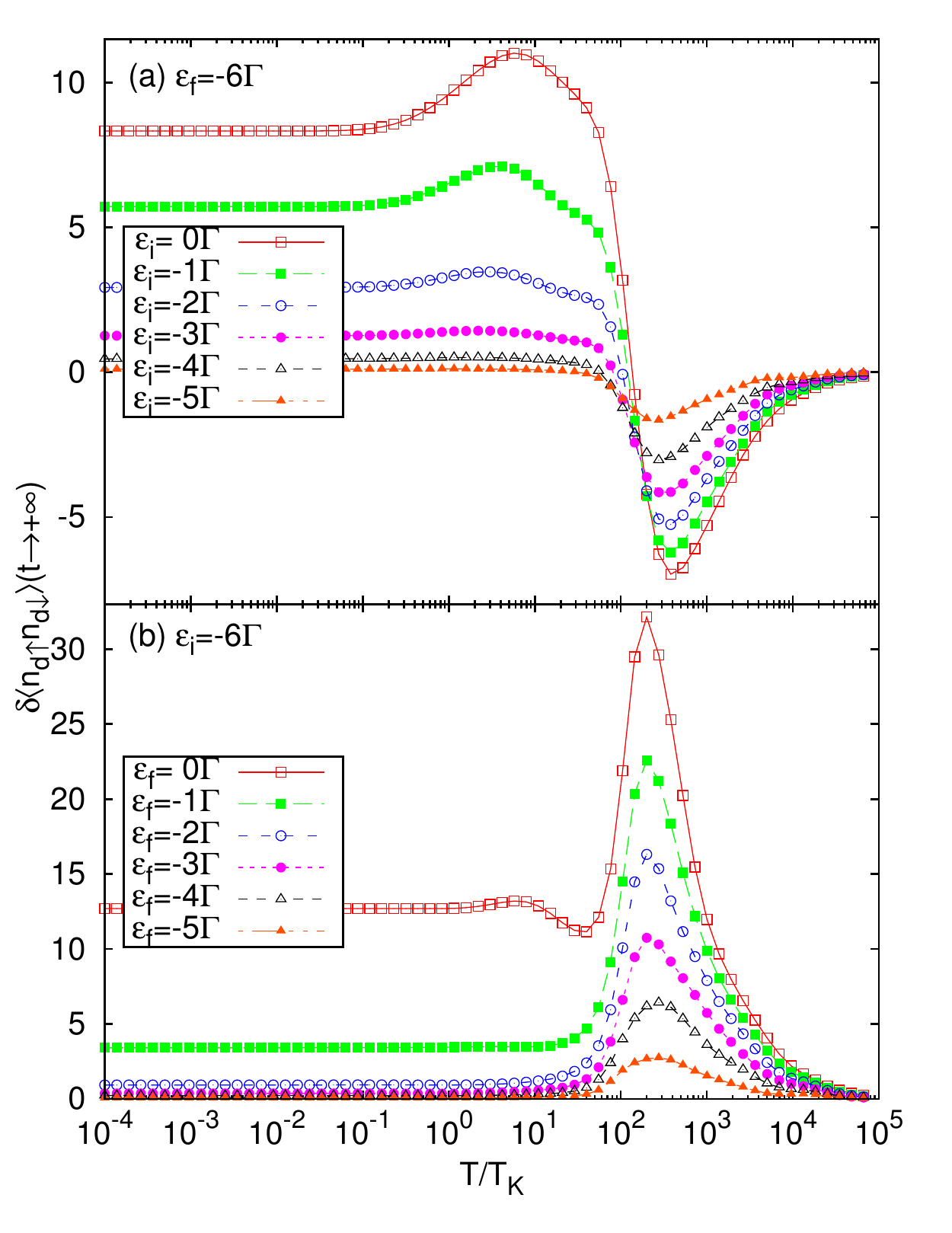}
\caption
{
  (Color online) 
  The (percentage) relative error of the double occupation number in the long time limit vs temperatures. $U_i=U_f=U=12\Gamma$ and $\Gamma=10^{-3}D$. 
  (a) Switching from the asymmetric to the symmetric Kondo system. 
  (b) Reverse switching, from the symmetric to the asymmetric Kondo system [the $\varepsilon_f$ are indicated and are the same as the $\varepsilon_i$ in (a)]. 
  $T_K\approx 2.0\times 10^{-5}D$ is the Kondo temperature of the symmetric model and calculations are for $\Lambda=1.6$ and no $z$ averaging.
}
\label{fig:figure11}
\end{figure}
In Fig.~\ref{fig:figure11}, we present the error in the double occupancy upon switching the system between the asymmetric case and the symmetric Kondo regime
while keeping the Coulomb interaction constant, i.e.  $\varepsilon_d(t)=\varepsilon_i\theta(-t)+\varepsilon_f\theta(t)$ with $U=U_i=U_f$. The analogous results for the occupation number  were presented  in Fig.~\ref{fig:Otinf}.
As for the error in the occupation number in Fig.~\ref{fig:Otinf}, we observe for that the error in the double occupancy exhibits extrema at approximately the same temperatures, namely at $T\approx 300T_K\approx U/2$ and $T\approx 7T_K=0.14\Gamma$.
The error increases with increasing quench size for both the switching from the asymmetric model to the symmetric Kondo regime [Fig.~\ref{fig:figure11}~(a)] and vice versa [Fig.~\ref{fig:figure11}~(b)], supporting the first trend discussed in Sec.~\ref{subsec:long-time-limit}.
The highest incoherent excitation in Fig.~\ref{fig:figure11}~(a), defined by Eq.~(\ref{eq:incoherent}), is $6\Gamma$, while in Fig.~\ref{fig:figure11}~(b), it varies from $7\Gamma$ up to $12\Gamma$. For the same sized quench in  Figs.~\ref{fig:figure11}~(a) and \ref{fig:figure11}(b), we see that
the error is always larger for  Fig.~\ref{fig:figure11}~(b), thus supporting the conclusion that this error depends on the magnitude of the highest incoherent excitation in the final state.
\subsection{Occupation number and double occupancy: switching  $U$ with $\varepsilon_d$ constant}
\begin{figure}[h]
\centering
\includegraphics[width=\linewidth,clip]{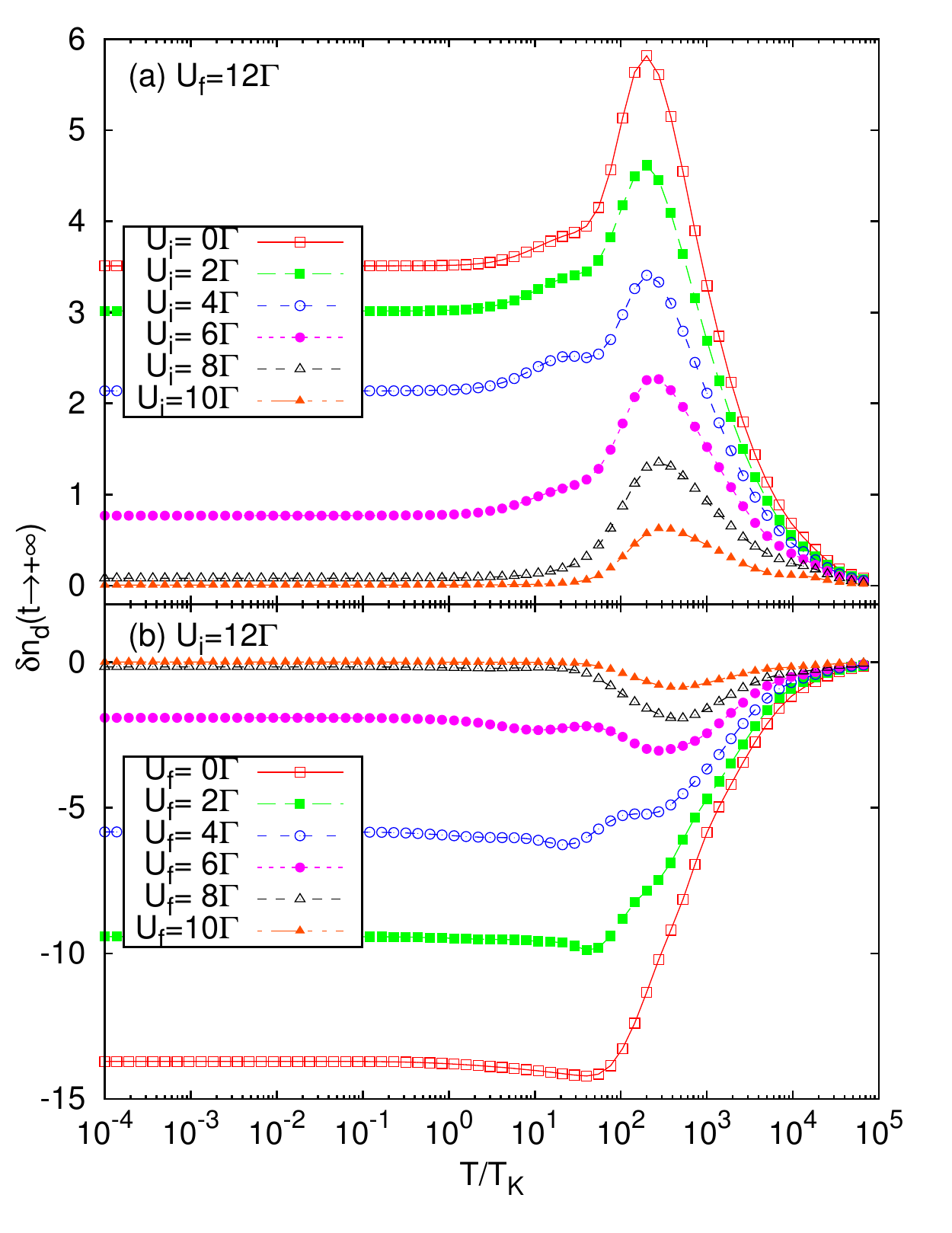}
\caption{
  (Color online) 
  The (percentage) relative error of the local level occupation number in the long time limit vs temperature. $\varepsilon_i=\varepsilon_f=-6\Gamma$ and $\Gamma=10^{-3}D$. 
  (a) Switching from the asymmetric model to the symmetric Kondo regime. 
  (b) Reverse switching, from the symmetric Kondo regime to the asymmetric model [the $\varepsilon_f$ are indicated and are the same as the $\varepsilon_i$ in (a)]. 
  $T_K\approx 2.0\times 10^{-5}D$ is the Kondo temperature of the symmetric model and calculations are for $\Lambda=1.6$ and no $z$ averaging.
}
\label{fig:figure12}
\end{figure}
\begin{figure}[h]
\centering
\includegraphics[width=\linewidth,clip]{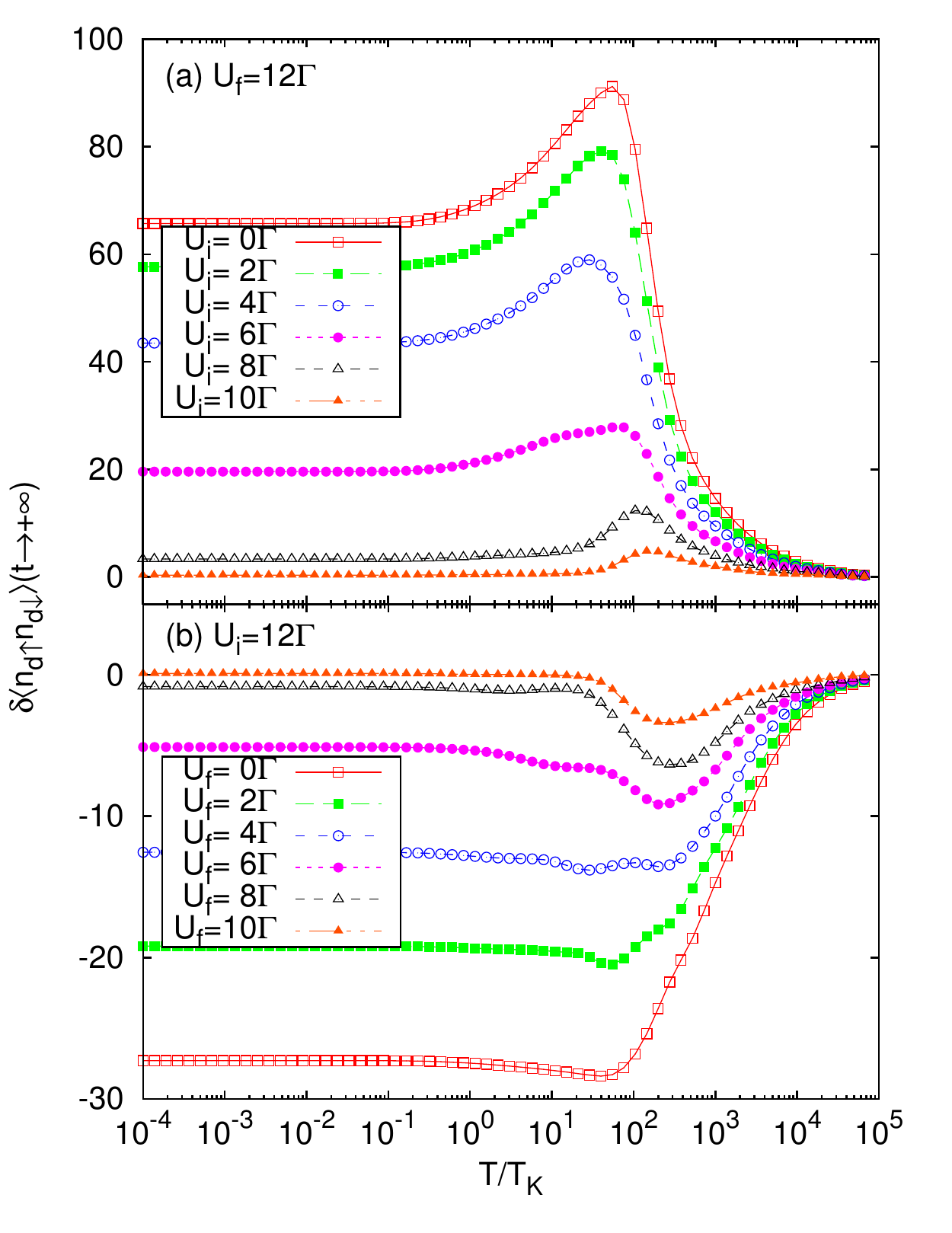}
\caption
{
  (Color online) 
  The (percentage) relative error of the double occupancy in the long time limit vs temperatures. 
  Parameters and switching in (a) and (b) are respectively the same as those in Fig.~\ref{fig:figure12}~(a) and (b).
}
\label{fig:figure13}
\end{figure}
In Sec.~\ref{subsec:long-time-limit} we discussed the error in the long time limit of the double occupancy for the case where both $U$ and $\varepsilon_d$ were switched at $t=0$, while maintaining particle-hole symmetry, i.e. for $\varepsilon_d(t)+U(t)/2=0$.
Here, we consider relaxing the last condition, and consider changing the Coulomb interaction $U(t)=U_i\theta(-t)+U_f\theta(t)$ while keeping $\varepsilon_d=\varepsilon_i = \varepsilon_f =-6\Gamma$ constant in both initial and final states. We either choose
$U_i=12\Gamma$ and switch $U_f$, or vice versa. Figure~\ref{fig:figure12} shows the error in the long time limit of the occupation number and Fig.~\ref{fig:figure13} the corresponding error for the double occupancy.
For both quantities, the error increases with increasing size of the quantum quench and the positions of the extrema in the error are consistent with those seen in previous quenches. 
The trend that the error increases with increasing value of the highest incoherent excitation in the final state [Eq.~(\ref{eq:incoherent})] can not be applied in this case as the value of this excitation is the same ($6\Gamma$) for both switchings and all cases. 
However, Fig.~\ref{fig:figure12} agrees with the trend in Fig.~\ref{fig:Otinf} that switching from the symmetric to the asymmetric model results in a larger error than vice versa, and Fig.~\ref{fig:figure13} agrees with the trend in Fig.~\ref{fig:figure6} 
that a larger error is expected upon switching from a less correlated to a more correlated system than vice versa.
\section{Generalized overlap matrix elements}
\label{sec:tau-overlap-matrices}
In this appendix, we prove that the generalized overlap matrix elements at an arbitrary time step are diagonal in the environment variables, and that they 
can be calculated recursively. 

We start with the simplest case for the first time step $\tilde{\tau}_1$
\begin{align}
\mathcal{S}^{m}_{r_{Q_2}s_i}(ee',-\tilde{\tau}_1)&={_{Q_2}}\langle rem|e^{-iH^{Q_1}{\tau}_1}|se'm\rangle{_i}\nonumber\\
&=\sum_{m_1l_1e_1}{_{Q_2}}\langle rem|l_1e_1m_1\rangle{_{Q_1}}{_{Q_1}}\langle l_1e_1m_1|e^{-iH^{Q_1}{\tau}_1}|se'm\rangle{_i}\nonumber.
\end{align}
Decomposing the matrix element into three terms, corresponding to $m_1>m$, $m_1=m$, and $m_1<m$, contributions, respectively, we have
\begin{align}
&\mathcal{S}^{m}_{r_{Q_2}s_i}(ee',-\tilde{\tau}_1)\nonumber\\
&=\mathcal{S}^{m++}_{r_{Q_2}s_i}(ee',-\tilde{\tau}_1)+\mathcal{S}^{m0}_{r_{Q_2}s_i}(ee',-\tilde{\tau}_1)+\mathcal{S}^{m--}_{r_{Q_2}s_i}(ee',-\tilde{\tau}_1),\nonumber\end{align}
with each term shown to be diagonal in the environment variables as follows
\begin{align}
&\mathcal{S}^{m++}_{r_{Q_2}s_i}(ee',-\tilde{\tau}_1)\nonumber\\
&=\sum_{m_1=m+1}^N\sum_{l_1e_1}{_{Q_2}}\langle rem|l_1e_1m_1\rangle{_{Q_1}}{_{Q_1}}\langle l_1e_1m_1|e^{-iH^{Q_1}{\tau}_1}|se'm\rangle{_i}\nonumber\\
&=\sum_{k}{_{Q_2}}\langle rem|kem\rangle{_{Q_1}}{_{Q_1}}\langle kem|e^{-iH^{Q_1}{\tau}_1}|se'm\rangle{_i}\nonumber\\
&=\sum_{k}S^m_{r_{Q_2}k_{Q_1}}e^{-iE^m_k{\tau}_1}S^m_{k_{Q_1}s_i}\delta_{ee'}=\mathcal{S}^{m++}_{r_{Q_2}s_i}(-\tilde{\tau}_1)\delta_{ee'}\label{eq:Stau1++}\end{align}
\begin{align}
&\mathcal{S}^{m0}_{r_{Q_2}s_i}(ee',-\tilde{\tau}_1)
=\sum_{l}{_{Q_2}}\langle rem|lem\rangle{_{Q_1}}{_{Q_1}}\langle lem|e^{-iH^{Q_1}{\tau}_1}|se'm\rangle{_i}\nonumber\\
&=\sum_{l}S^m_{r_{Q_2}l_{Q_1}}e^{-iE^m_l{\tau}_1}S^m_{l_{Q_1}s_i}\delta_{ee'}=\mathcal{S}^{m0}_{r_{Q_2}s_i}(-\tilde{\tau}_1)\delta_{ee'}\label{eq:Stau10}\\
&\mathcal{S}^{m--}_{r_{Q_2}s_i}(ee',-\tilde{\tau}_1)\nonumber\\
&=\sum_{m_1=m_0}^{m-1}\sum_{l_1e_1}{_{Q_2}}\langle rem|l_1e_1m_1\rangle{_{Q_1}}{_{Q_1}}\langle l_1e_1m_1|e^{-iH^{Q_1}{\tau}_1}|se'm\rangle{_i}\nonumber\\
&=\sum_{m_1=m_0}^{m-1}\sum_{l_1e_1}{_{Q_2}}\langle rem|(1^{Q_2+}_{m_1}+1^{Q_2-}_{m_1})|l_1e_1m_1\rangle{_{Q_1}}\nonumber\\
&\hspace{4em}\times{_{Q_1}}\langle l_1e_1m_1|e^{-iH^{Q_1}{\tau}_1}(1^{i+}_{m_1}+1^{i-}_{m_1})|se'm\rangle{_i}\nonumber\\
&=\sum_{m_1=m_0}^{m-1}\sum_{kk'e_1e'_1}{_{Q_2}}\langle rem|ke_1m_1\rangle{_{Q_2}}\sum_{l_1}{_{Q_2}}\langle ke_1m_1|l_1e_1m_1\rangle{_{Q_1}}\nonumber\\
&\hspace{2em}\times{_{Q_1}}\langle l_1e_1m_1|e^{-iH^{Q_1}{\tau}_1}|k'e'_1m_1\rangle{_{i}}{_{i}}\langle k'e'_1m_1|se'm\rangle{_i}\nonumber\\
&=\sum_{m_1=m_0}^{m-1}\sum_{kk'e_1}{_{Q_2}}\langle rem|ke_1m_1\rangle{_{Q_2}}\nonumber\\
&\hspace{2em}\times\Bigg[\sum_{l_1}S^{m_1}_{k_{Q_2}l_{1Q_1}}e^{-iE^{m_1}_{l_1}{\tau}_1}S^{m_1}_{l_{1Q_1}k'_i}\Bigg]{_{i}}\langle k'e_1m_1|se'm\rangle{_i}\nonumber.\end{align}
Using the definition of $\mathcal{S}^{m0}_{r_{Q_2}s_i}(-\tilde{\tau}_1)$ in Eq.~(\ref{eq:Stau10}), we have
\begin{align}
&\mathcal{S}^{m--}_{r_{Q_2}s_i}(ee',-\tilde{\tau}_1)\nonumber\\
&=\sum_{m_1=m_0}^{m-1}\sum_{kk'e_1}{_{Q_2}}\langle rem|ke_1m_1\rangle{_{Q_2}}\Big[\mathcal{S}^{m_10}_{k_{Q_2}k'_i}(-\tilde{\tau}_1)\Big]{_{i}}\langle k'e_1m_1|se'm\rangle{_i}\nonumber.\end{align}
Noticing that the structure of the above equation is similar to that of Eq.~(\ref{eq:rho--}), we derive the recursion relation of $\mathcal{S}^{m--}_{r_{Q_2}s_i}(ee',-\tilde{\tau}_1)$ following the procedure for $\rho^{--}$ in sec.~\ref{sec:section-3}, thus we have
\begin{align}
&\mathcal{S}^{m--}_{r_{Q_2}s_i}(ee',-\tilde{\tau}_1)\nonumber\\
&=\sum_{\alpha_m}\sum_{kk'}A^{\alpha_m\dagger}_{rk}\Big[\mathcal{S}^{(m-1)0}_{k_{Q_2}k'_i}(-\tilde{\tau}_1)+\mathcal{S}^{(m-1)--}_{k_{Q_2}k'_i}(-\tilde{\tau}_1)\Big]A^{\alpha_m}_{k's}\delta_{ee'}\nonumber\\
&=\mathcal{S}^{m--}_{r_{Q_2}s_i}(-\tilde{\tau}_1)\delta_{ee'}\quad {\text with}\quad \mathcal{S}^{m_0--}_{r_{Q_2}s_i}(-\tilde{\tau}_1)=0.\label{eq:Stau1--}\end{align}
Since each term of the generalized overlap matrix element is diagonal in the environment variables, we define 
\begin{align}
\mathcal{S}^m_{r_{Q_2}s_i}(ee',-\tilde{\tau}_1)&=\mathcal{S}^m_{r_{Q_2}s_i}(-\tilde{\tau}_1)\delta_{ee'}
\\ 
\text{with\quad } \mathcal{S}^m_{r_{Q_2}s_i}(-\tilde{\tau}_1)&=\mathcal{S}^{m++}_{r_{Q_2}s_i}(-\tilde{\tau}_1)+\mathcal{S}^{m0}_{r_{Q_2}s_i}(-\tilde{\tau}_1)+\mathcal{S}^{m--}_{r_{Q_2}s_i}(-\tilde{\tau}_1)\nonumber.
\end{align} 

In a similar way, we consider the generalized overlap matrix element at the next time step $\tilde{\tau}_2$, 
\begin{align}
&\mathcal{S}^{m}_{r_{Q_3}s_i}(ee',-\tilde{\tau}_2)\nonumber\\&={_{Q_3}}\langle rem|e^{-iH^{Q_2}{\tau}_2}e^{-iH^{Q_1}{\tau}_1}|se'm\rangle{_i}\nonumber\\
&=\sum_{m_2l_2e_2}{_{Q_3}}\langle rem|l_2e_2m_2\rangle{_{Q_2}}{_{Q_2}}\langle l_2e_2m_2|e^{-iH^{Q_2}{\tau}_2}e^{-iH^{Q_1}{\tau}_1}|se'm\rangle{_i}\nonumber\\
&=\mathcal{S}^{m++}_{r_{Q_3}s_i}(ee',-\tilde{\tau}_2)+\mathcal{S}^{m0}_{r_{Q_3}s_i}(ee',-\tilde{\tau}_2)+\mathcal{S}^{m--}_{r_{Q_3}s_i}(ee',-\tilde{\tau}_2).\end{align}
The three terms above correspond to $m_2>m$, $m_2=m$, and $m_2<m$, respectively, and are shown to be diagonal in the environment variables as follows,
\begin{align}
&\mathcal{S}^{m++}_{r_{Q_3}s_i}(ee',-\tilde{\tau}_2)\nonumber\\
&=\sum_{m_2=m+1}^N\sum_{l_2e_2}{_{Q_3}}\langle rem|l_2e_2m_2\rangle{_{Q_2}}{_{Q_2}}\langle l_2e_2m_2|e^{-iH^{Q_2}{\tau}_2}e^{-iH^{Q_1}{\tau}_1}|se'm\rangle{_i}\nonumber\\
&=\sum_{k}{_{Q_3}}\langle rem|kem\rangle{_{Q_2}}{_{Q_2}}\langle kem|e^{-iH^{Q_2}{\tau}_2}e^{-iH^{Q_1}{\tau}_1}|sem\rangle{_i}\nonumber\\
&=\sum_{k}{_{Q_3}}\langle rem|kem\rangle{_{Q_2}}e^{-iE^m_k{\tau}_2}{_{Q_2}}\langle kem|e^{-iH^{Q_1}{\tau}_1}|se'm\rangle{_i}\nonumber\\
&=\sum_{k}S^m_{r_{Q_3}k_{Q_2}}e^{-iE^m_k{\tau}_2}\mathcal{S}^m_{k_{Q_2}s_i}(-\tilde{\tau}_1)\delta_{ee'}=\mathcal{S}^{m++}_{r_{Q_3}s_i}(-\tilde{\tau}_2)\delta_{ee'}\end{align}
\begin{align}
&\mathcal{S}^{m0}_{r_{Q_3}s_i}(ee',-\tilde{\tau}_2)\nonumber\\
&=\sum_{l}{_{Q_3}}\langle rem|lem\rangle{_{Q_2}}{_{Q_2}}\langle lem|e^{-iH^{Q_2}{\tau}_2}e^{-iH^{Q_1}{\tau}_1}|se'm\rangle{_i}\nonumber\\
&=\sum_{l}{_{Q_3}}\langle rem|lem\rangle{_{Q_2}}e^{-iE^m_l{\tau}_2}{_{Q_2}}\langle lem|e^{-iH^{Q_1}{\tau}_1}|se'm\rangle{_i}\nonumber\\
&=\sum_{l}S^m_{r_{Q_3}l_{Q_2}}e^{-iE^m_l{\tau}_2}\mathcal{S}^m_{l_{Q_2}s_i}(-\tilde{\tau}_1)\delta_{ee'}=\mathcal{S}^{m0}_{r_{Q_3}s_i}(-\tilde{\tau}_2)\delta_{ee'}\end{align}
\begin{align}
&\mathcal{S}^{m--}_{r_{Q_3}s_i}(ee',-\tilde{\tau}_2)\nonumber\\
&=\sum_{m_2=m_0}^{m-1}\sum_{l_2e_2}{_{Q_3}}\langle rem|l_2e_2m_2\rangle{_{Q_2}}{_{Q_2}}\langle l_2e_2m_2|e^{-iH^{Q_2}{\tau}_2}e^{-iH^{Q_1}{\tau}_1}|se'm\rangle{_i}\nonumber\\
&=\sum_{m_2=m_0}^{m-1}\sum_{kk'e_2e'_2}{_{Q_3}}\langle rem|ke_2m_2\rangle{_{Q_3}}\sum_{l_2}{_{Q_3}}\langle ke_2m_2|l_2e_2m_2\rangle{_{Q_2}}\nonumber\\
&\hspace{5em}\times{_{Q_2}}\langle l_2e_2m_2|e^{-iH^{Q_2}{\tau}_2}e^{-iH^{Q_1}{\tau}_1}|k'e'_2m_2\rangle{_{i}}{_{i}}\langle k'e'_2m_2|se'm\rangle{_i}\nonumber\\
&=\sum_{m_2=m_0}^{m-1}\sum_{kk'e_2}{_{Q_3}}\langle rem|ke_2m_2\rangle{_{Q_3}}\Bigg[\sum_{l_2}S^{m_2}_{k_{Q_3}l_{2Q_2}}e^{-iE^{m_2}_{l_2}{\tau}_2}\mathcal{S}_{l_{2Q_2}k'_i}(m_2,-\tilde{\tau}_1)\Bigg]\nonumber\\
&\hspace{12em}\times{_{i}}\langle k'e_2m_2|se'm\rangle{_i}\nonumber\\
&=\sum_{m_2=m_0}^{m-1}\sum_{kk'e_2}{_{Q_3}}\langle rem|ke_2m_2\rangle{_{Q_3}}\Big[\mathcal{S}^{m_20}_{k_{Q_3}k'_i}(-\tilde{\tau}_2)\Big]{_{i}}\langle k'e_2m_2|se'm\rangle{_i}\nonumber.\end{align}
Again, noticing that the structure of the last expression above is similar to that in Eq.~(\ref{eq:rho--}), we derive the following recursion relation
\begin{align}
&\mathcal{S}^{m--}_{r_{Q_3}s_i}(ee',-\tilde{\tau}_2)\nonumber\\
&=\sum_{\alpha_m}\sum_{kk'}A^{\alpha_m\dagger}_{rk}\Big[\mathcal{S}^{(m-1)0}_{k_{Q_3}k'_i}(-\tilde{\tau}_2)+\mathcal{S}^{(m-1)--}_{k_{Q_3}k'_i}(-\tilde{\tau}_2)\Big]A^{\alpha_m}_{rs}\delta_{ee'}\nonumber\\
&=\mathcal{S}^{m--}_{r_{Q_3}s_i}(-\tilde{\tau}_2)\delta_{ee'} \quad \text{with} \quad \mathcal{S}^{m_0--}_{r_{Q_3}s_i}(-\tilde{\tau}_2)=0. 
\end{align}
Since each term of the overlap matrix element at $\tilde{\tau}_2$ is also diagonal in the environment variables, we have
\begin{align}
\mathcal{S}^m_{r_{Q_3}s_i}(ee',-\tilde{\tau}_2)&=\mathcal{S}^m_{r_{Q_3}s_i}(-\tilde{\tau}_2)\delta_{ee'}
\\ 
\text{with\quad } \mathcal{S}^m_{r_{Q_3}s_i}(-\tilde{\tau}_2)&=\mathcal{S}^{m++}_{r_{Q_3}s_i}(-\tilde{\tau}_2)+\mathcal{S}^{m0}_{r_{Q_3}s_i}(-\tilde{\tau}_2)+\mathcal{S}^{m--}_{r_{Q_3}s_i}(-\tilde{\tau}_2)\nonumber.
\end{align} 
Consequently, the generalized overlap matrix elements at an arbitrary time step $\tilde{\tau}_{p+1}>\tilde{\tau}_1$ can decomposed as follows
\begin{align}
&\mathcal{S}^{m}_{r_{Q_{p+1}}s_i}(ee',-\tilde{\tau}_{p})\nonumber\\&={_{Q_{p+1}}}\langle rem|e^{-iH^{Q_p}{\tau}_p}...e^{-iH^{Q_1}{\tau}_1}|se'm\rangle{_i}\nonumber\\
&=\sum_{m_pl_pe_p}{_{Q_{p+1}}}\langle rem|l_pe_pm_p\rangle{_{Q_p}}{_{Q_p}}\langle l_pe_pm_p|e^{-iH^{Q_p}{\tau}_2}...e^{-iH^{Q_1}{\tau}_1}|se'm\rangle{_i}\nonumber\\
&=\mathcal{S}^{m++}_{r_{Q_{p+1}}s_i}(ee',-\tilde{\tau}_p)+\mathcal{S}^{m0}_{r_{Q_{p+1}}s_i}(ee',-\tilde{\tau}_p)+\mathcal{S}^{m--}_{r_{Q_{p+1}}s_i}(ee',-\tilde{\tau}_p)
\nonumber.\end{align}
Again, the three terms above correspond to the $m>m_p$, $m=m_p$, and $m<m_p$ contributions, and satisfy the general rule
\begin{align}
\mathcal{S}^{m++}_{r_{Q_{p+1}}s_i}(ee',-\tilde{\tau}_p)&=\sum_{k}S^m_{r_{Q_{p+1}}k_{Q_p}}e^{-iE^m_k{\tau}_p}\mathcal{S}^m_{k_{Q_p}s_i}(-\tilde{\tau}_{p-1})\delta_{ee'}\nonumber\\
&=\mathcal{S}^{m++}_{r_{Q_{p+1}}s_i}(-\tilde{\tau}_p)\delta_{ee'}\label{eq:Staun++}\end{align}
\begin{align}
\mathcal{S}^{m0}_{r_{Q_{p+1}}s_i}(ee',-\tilde{\tau}_p)&=\sum_{l}S^m_{r_{Q_{p+1}}l_{Q_p}}e^{-iE^m_l{\tau}_p}\mathcal{S}^m_{l_{Q_p}s_i}(-\tilde{\tau}_{p-1})\delta_{ee'}\nonumber\\
&=\mathcal{S}^{m0}_{r_{Q_{p+1}}s_i}(-\tilde{\tau}_p)\delta_{ee'}\label{eq:Staun0}
\end{align}
\begin{align}
&\mathcal{S}^{m--}_{r_{Q_{p+1}}s_i}(ee',-\tilde{\tau}_p)\nonumber\\
&=\sum_{\alpha_m}\sum_{kk'}A^{\alpha_m\dagger}_{rk}\Big[\mathcal{S}^{(m-1)0}_{k_{Q_{p+1}}k'_i}(-\tilde{\tau}_{p})+\mathcal{S}^{(m-1)--}_{k_{Q_{p+1}}k'_i}(-\tilde{\tau}_{p})\Big]A^{\alpha_m}_{k's}\delta_{ee'}\nonumber\\
&=\mathcal{S}^{m--}_{r_{Q_{p+1}}s_i}(-\tilde{\tau}_p)\delta_{ee'}\quad\text{with\quad } \mathcal{S}^{m_0--}_{r_{Q_{p+1}}s_i}(-\tilde{\tau}_p)=0\label{eq:Staun--}.\end{align}
Since each term of $\mathcal{S}^m_{r_{f}s_i}(ee',-\tilde{\tau}_p)$ is diagonal in the environment variables, we have
\begin{align}
\mathcal{S}^m_{r_{Q_{p+1}}s_i}(ee',-\tilde{\tau}_p)&=\mathcal{S}^m_{r_{Q_{p+1}}s_i}(-\tilde{\tau}_p)\delta_{ee'}\\ 
\text{with\quad}
\mathcal{S}^m_{r_{Q_{p+1}}s_i}(-\tilde{\tau}_p)&=\mathcal{S}^{m++}_{r_{Q_{p+1}}s_i}(-\tilde{\tau}_p)+\mathcal{S}^{m0}_{r_{Q_{p+1}}s_i}(-\tilde{\tau}_p)+\mathcal{S}^{m--}_{r_{Q_{p+1}}s_i}(-\tilde{\tau}_p)\nonumber.
\end{align}

In general, we showed that the generalized  overlap matrix elements at an arbitrary time step are diagonal in the environment variables. Moreover, one can use Eqs.~(\ref{eq:Stau1++})-(\ref{eq:Stau1--}) to calculate each term of the generalized overlap matrix element at $\tilde{\tau}_1$, and Eqs.~(\ref{eq:Staun++})-(\ref{eq:Staun--}) to calculate those at $\tilde{\tau}_p>\tilde{\tau}_1$.
\bibliography{noneq-nrg}
\end{document}